\documentclass[a4paper,11pt]{article}

\usepackage{jheppub}
\usepackage[T1]{fontenc}
\usepackage{graphicx}% Include figure files
\usepackage{hyperref}
\usepackage{amsmath}
\usepackage{epsfig}
\usepackage{subfig}
\usepackage{xcolor}
\usepackage{natbib}
\usepackage{bm}% bold math

\def\be{\begin{equation}}
\def\ee{\end{equation}}
\def\bea{\begin{eqnarray}}
\def\eea{\end{eqnarray}}
\def\NO{\nonumber}
\def\gev{\mathrm{~GeV}}

\def\md{\mathrm{d}}

\title{Statistical analysis of the azimuthal asymmetry in the $J/\psi$ leptoproduction in unpolarized $ep$ collisions}% Force line breaks with \\

\author[a]{Hong-Fei Zhang,}
\author[b,1]{Wen-Long Sang\note{Corresponding Author.}}
\author[c]{Yu-Peng Yan}
\affiliation[a]{College of Big Data Statistics, Guizhou University of Finance and Economics, Guiyang, 550025, China}
\affiliation[b]{School of Physical Science and Technology, Southwest University, Chongqing, 400700, China}
\affiliation[c]{School of Physics and Center of Excellence in High Energy Physics $\mathrm{\&}$ Astrophysics,
Suranaree University of Technology, Nakhon Ratchasima 30000, Thailand}
\emailAdd{hfzhang@ihep.ac.cn}
\emailAdd{wlsang@ihep.ac.cn}
\emailAdd{yupeng@sut.ac.th}

\abstract{
In this paper, we study the azimuthal asymmetry in the $J/\psi$ leptoproduction in unpolarized $ep$ collisions.
There are two independent azimuthal asymmetry modulations, namely $\mathrm{cos}(\psi)$ and $\mathrm{cos}(2\psi)$,
where $\psi$ is the azimuthal angle of the lepton scattering plane with respect to the hadron-interacting plane.
We calculate the two modulations as functions of four kinematic variables,
and find that they provide a very good laboratory to distinguish several models describing the heavy quarkonium production,
including the color-singlet (CS) model, the nonrelativistic QCD (NRQCD) associated with the $^1S_0^{[8]}$ dominance picture,
and the NRQCD in which the values of all the three color-octet (CO) long-distance matrix elements are of the same order.
In order to make definite conclusions, we restrict our calculation in a specific kinematic region,
where the CS and CO mechanisms can be distinguished by scrutinizing the values of the $\mathrm{cos}(\psi)$ modulation,
while the $^1S_0^{[8]}$ dominance picture can be tested by measuring the values of the $\mathrm{cos}(2\psi)$ modulation.
Calculating their values and carrying out a meticulous statistical analysis,
we find that at an integrated luminosity $\mathcal{L}=1000\mathrm{pb}^{-1}$,
the statistical uncertainties of the two quantities are small enough to tell the three models apart.
When this experiment is implemented at the future $ep$ colliders such as the EIC,
crucial information for the $J/\psi$ production mechanism might be discovered.
}
\begin{document}

\maketitle

\bibliographystyle{JHEP}

\section{Introduction\label{sec:introduction}}

It has not been realized that the transverse polarization and transverse momentum effects in nucleons could be significant
until the first measurement of inclusive pion hadroproduction was carried out at Argonne synchrotron~\cite{Dick:1975ty, Klem:1976ui, Dragoset:1978gg},
the results of which were later confirmed by Fermilab~\cite{Bunce:1976yb} at a slightly higher colliding energy.
In order to understand the unexpected large transverse spin asymmetries observed in those experiments,
various theoretical mechanisms and new structure functions were proposed,
among which are there the well-known Sivers effect~\cite{Sivers:1989cc, Sivers:1990fh} along with a new function, the Sivers function,
describing the azimuthal asymmetry of unpolarized quarks in polarized hadrons,
and Collins effect~\cite{Collins:1992kk, Collins:1993kq},
considering the different fragmenting probabilities of transversely polarized partons to hadrons with transverse momentum along different directions.
Since this century, the transverse spin effects have been studied experimentally in semi-inclusive deep-inelastic scattering (SIDIS).
In 2004, HERMES~\cite{Airapetian:2004tw} and COMPASS~\cite{Alexakhin:2005iw, Ageev:2006da} published their measurements
of the single-spin asymmetry in the collisions of leptons off polarized proton and deuteron targets.
One of the main advantages of SIDIS is that the transverse spin and transverse momentum effects are not mixed, as in hadroproduction;
they result in different azimuthal asymmetry modulations.
Another interesting feature of SIDIS is that, even with unpolarized targets,
there also exist two types of azimuthal asymmetry modulations, namely the $\mathrm{cos}(\psi)$ and $\mathrm{cos}(2\psi)$ modulations,
where $\psi$ is the azimuthal angle for the observed final-state hadron with respect to the virtual-photon-target beam.
A few years ago, HERMES~\cite{Airapetian:2012yg} and COMPASS~\cite{Adolph:2014pwc} collaborations
presented the data for the azimuthal distributions of hadrons produced in deep inelastic scattering
off unpolarized targets and found nonzero $\mathrm{cos}(\psi)$ and $\mathrm{cos}(2\psi)$ azimuthal asymmetry modulations. 
These modulations arise as long as the momenta of the final-state hadrons have transverse components,
which may originate from either the intrinsic transverse motion of partons inside the targets,
or the hard emission of the final-state hadrons.
The former one was first studied by Cahn~\cite{Cahn:1978se, Cahn:1989yf} and thus is called the Cahn effect.
Cahn effect dominates in low $p_t$ region, while in high $p_t$ region,
the real emission plays a more important role.

The azimuthal asymmetries in unpolarized SIDIS not only provide essential information for the transverse insight of nucleons,
but also open a window for the study of hadron production mechanisms, e.g. the heavy quarkonium production mechanism.
Although heavy quarkonia have very simple structure, their production mechanism is still unknown.
Nonrelativistic QCD (NRQCD)~\cite{Bodwin:1994jh},
which separates the perturbatively calculable short-distance coefficients (SDCs) from the process-independent long-distance matrix elements (LDMEs),
is one of the most successful theories describing the quarkonium productions and decays.
However it confronted a lot of difficulties and challenges in the past two decades.
The most famous one of those is the $J/\psi$ polarization puzzle.
In the last decade, many processes have been calculated up to QCD next-to-leading order (NLO), including the $J/\psi$
\cite{Zhang:2005cha, Zhang:2006ay, Gong:2007db, Gong:2008ce, Ma:2008gq, Gong:2009kp, Gong:2009ng, Zhang:2009ym, Wang:2011qg, Feng:2017bdu, Jiang:2018wmv}
and the $\eta_c$~\cite{Gong:2016jiq} production in $e^+e^-$ annihilation,
the $J/\psi$ photoproduction in $e^+e^-$~\cite{Klasen:2004tz} and $ep$
\cite{Kramer:1995nb, Maltoni:1997pt, Artoisenet:2009xh, Chang:2009uj, Li:2009fd, Butenschoen:2009zy, Butenschoen:2011ks, Bodwin:2015yma} collisions,
the $\psi$~\cite{Campbell:2007ws, Gong:2008sn, Gong:2008hk, Gong:2008ft, Ma:2010yw, Butenschoen:2010rq, Ma:2010jj, Butenschoen:2011yh, Butenschoen:2012px,
Chao:2012iv, Gong:2012ug, Gong:2012ah, Lansberg:2013qka, Li:2014ava, Bodwin:2014gia, Lansberg:2014swa, Shao:2014yta, Sun:2015pia, Bodwin:2015iua, Feng:2018cai},
$\eta_c$~\cite{Butenschoen:2014dra, Han:2014jya, Zhang:2014ybe, Lansberg:2017ozx, Feng:2019zmn},
and $\chi_c$~\cite{Ma:2010vd, Li:2011yc, Shao:2012fs, Shao:2014fca, Jia:2014jfa} hadroproduction,
and the $J/\psi$ production in deep-inelastic scattering (DIS)~\cite{Sun:2017wxk}.
As a noteworthy progress, the measurement of the $\eta_c$ production~\cite{Barsuk:2012ic, Aaij:2014bga} stimulated a lot of theoretical works
\cite{Butenschoen:2014dra, Han:2014jya, Zhang:2014ybe, Gong:2016jiq, Lansberg:2017ozx, Feng:2019zmn, Zhang:2019wxo}.
Unfortunately, neither a set of globally working LDMEs has been found,
nor the importance of the color-octet (CO) mechanism has been demonstrated.
We refer the readers to Reference~\cite{Lansberg:2017ozx} as a state-of-the-art review.
Azimuthal asymmetry in the $J/\psi$ production in DIS can serve as an alternative test of NRQCD.
As we will see later, in some specific kinematic regions,
the azimuthal asymmetry can distinguish different production channels.
Based on statistical analysis, we propose a practical strategy to verify the CO mechanism and fix the corresponding LDMEs.

The rest of this paper is organized as follows.
The analytical framework of our calculation is discussed in Section~\ref{sec:framework}.
In Section~\ref{sec:numerical}, we present our numerical results,
according to which, the statistical analysis is given in the same section.
Section~\ref{sec:summary} is a concluding remark.

\section{The azimuthal asymmetry in the $J/\psi$ leptoproduction\label{sec:framework}}

\subsection{The calculation of the cross sections}

\begin{figure}
\includegraphics[scale=0.5]{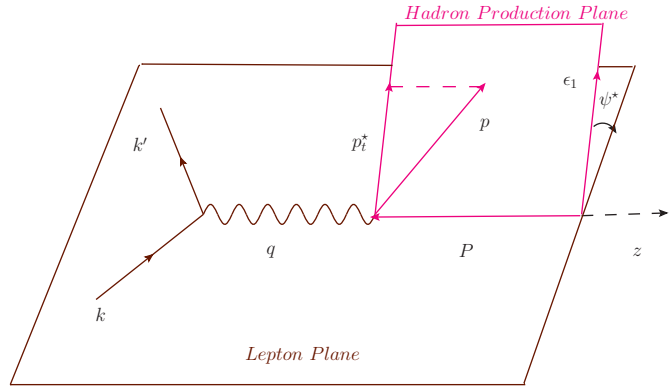}
\centering
\caption{\label{fig:process}
Illustration of the process.
}
\end{figure}

In electron-proton collisions, 
\bea
e(k)+p(P)\rightarrow e(k')+J/\psi(p)+X(p_X) \label{eqn:process},
\eea
which, in general, can be reduced into the virtual-photon-proton fusion process as
\bea
\gamma^\star(q)+p(P)\rightarrow J/\psi(p)+X(p_X). \label{eqn:vprocess}
\eea
Here, $X$ represents the proton remnant.
$k$, $P$, $q$, $k'$, $p$, and $p_X$ are the momenta of the initial-state electron, proton, and virtual photon,
the final-state electron, the $J/\psi$, and the proton remnant.
Usually, we use the following invariants to describe the kinematics of this process,
\bea
Q^2=-q^2,~~~~~~W^2=(P+q)^2,~~~~~~z=\frac{P\cdot p}{P\cdot q}, \label{eqn:qwz}
\eea
the first two of which can be replaced by the following two dimensionless invariants,
\bea
x=\frac{Q^2}{2P\cdot q},~~~~~~y=\frac{2P\cdot q}{(P+k)^2}, \label{eqn:xy}
\eea
as long as the invariant colliding energy squared, $S=(P+k)^2$, is given.
Here, $x$ is the well-known Bjorken-$x$, and $z$ is named the elasticity coefficient.

To study the azimuthal effects,
we adopt such frames in which the virtual photon and the proton beams are along the $z$ and anti-$z$ directions, respectively,
as illustrated in Figure~\ref{fig:process}.
The azimuthal angle of the final-state $J/\psi$ is then defined with respect to the virtual-photon-proton beams.
In such frames, the physical quantities are always labelled by a superscript $\star$,
in order to distinguish from those in laboratory frames.
Since the laboratory frame is not concerned in this paper, we just omit the superscript $\star$.
In addition to the invariants, $x$, $y$, and $z$ (or equivalently, $Q^2$, $W^2$, and $z$),
we need two more quantities,
the transverse momentum of the final-state $J/\psi$, $p_t$, and its azimuthal angle, $\psi$,
to sufficiently describe the kinematics of the $J/\psi$ leptoproduction process.
We define two dimensionless quantities,
\bea
\xi=\frac{p_t^2}{S},~~~~~~\eta=\frac{M^2}{S}, \label{eqn:xieta}
\eea
where $M$ is the $J/\psi$ mass,
so that the SIDIS process can be described by six dimensionless variables, $x$, $y$, $z$, $\xi$, $\eta$, and $\psi$.
Having all these kinematic variables,
we can write down the hadron production cross section in unpolarized SIDIS as (see e.g. Reference~\cite{Zhang:2017dia})
\bea
\md\sigma=\frac{\alpha}{128\pi^4}l_{\mu\nu}W_h^{\mu\nu}\frac{\md x}{x}\md y\frac{\md z}{z}\md\xi\md\psi, \label{eqn:cs}
\eea
where $\alpha$ is the electromagnetic fine-structure constant,
$W_h^{\mu\nu}$ is the hadronic tensor integrated over the phase-space of the final-state hadrons other than the $J/\psi$,
and the normalized leptonic tensor, $l_{\mu\nu}$, is defined as
\bea
l_{\mu\nu}=(-g_{\mu\nu}-\frac{q_{\mu}q_{\nu}}{Q^2})+\frac{(2k-q)_{\mu}(2k-q)_{\nu}}{Q^2}. \label{eqn:luv}
\eea
If only one final-state hadron is observed, the leptonic tensor, $l_{\mu\nu}$,
can be decomposed into the linear combination of four simpler tensors,
factorizing all the azimuthally dependent terms into the corresponding coefficients as
\bea
l^{\mu\nu}=A_g(-g^{\mu\nu}-\frac{q^\mu q^\nu}{Q^2})+A_L\epsilon_L^\mu\epsilon_L^\nu
+A_{LT}(\epsilon_L^\mu\epsilon_T^\nu+\epsilon_T^\mu\epsilon_L^\nu)+A_T\epsilon_T^\mu\epsilon_T^\nu, \label{eqn:fluv}
\eea
where
\bea
&&\epsilon_L=\frac{1}{Q}(q+\frac{Q^2}{P\cdot q}P), \NO \\
&&\epsilon_T=\frac{1}{p_t}(p-\rho P-zq), \label{eqn:nv}
\eea
and
\bea
&&A_g=1+\frac{2(1-y)}{y^2}-\frac{2(1-y)}{y^2}\mathrm{cos}(2\psi), \NO \\
&&A_L=1+\frac{6(1-y)}{y^2}-\frac{2(1-y)}{y^2}\mathrm{cos}(2\psi), \NO \\
&&A_{LT}=\frac{2(2-y)}{y^2}\sqrt{1-y}\mathrm{cos}(\psi), \NO \\
&&A_T=\frac{4(1-y)}{y^2}\mathrm{cos}(2\psi). \label{eqn:A}
\eea
Here, $\rho$ is defined as
\bea
\rho=\frac{p\cdot q+zQ^2}{P\cdot q}=\frac{(\xi+\eta)/z+xyz}{y}, \label{eqn:rho}
\eea
where the proton mass has been neglected.
Substituting Equation~(\ref{eqn:nv}) into Equation~(\ref{eqn:fluv}) and taking into account the current-conserving equation,
\bea
q_\mu W_h^{\mu\nu}\equiv 0, \label{eqn:curcons}
\eea
we can rewrite the normalized leptonic tensor in a form that is more convenient for computation as
\bea
l^{\mu\nu}=C_1(-g^{\mu\nu})+C_2\frac{P^\mu P^\nu}{S}+C_3\frac{P^\mu p^\nu+p^\mu P^\nu}{2S}+C_4\frac{p^\mu p^\nu}{S}, \label{eqn:cluv}
\eea
where
\bea
&&C_1=A_g, \NO \\
&&C_2=\frac{4x}{y}A_L-4\rho\sqrt{\frac{x}{y\xi}}A_{LT}+\frac{\rho^2}{\xi}A_T, \NO \\
&&C_3=4\sqrt{\frac{x}{y\xi}}A_{LT}-\frac{2}{\xi}A_T, \NO \\
&&C_4=\frac{1}{\xi}A_T. \label{eqn:cn}
\eea
If we define
\bea
\beta=\frac{\rho}{2}\sqrt{\frac{y}{x\xi}}=\frac{\xi+\eta+xyz^2}{2z}\sqrt{\frac{1}{xy\xi}}, \label{eqn:beta}
\eea
these $C_i$'s can be expressed in a more compact form as
\bea
&&C_1=A_g, \NO \\
&&C_2=\frac{4x}{y}(A_L-2\beta A_{LT}+\beta^2A_T) \NO \\
&&C_3=4\sqrt{\frac{x}{y\xi}}(A_{LT}+\beta A_T) \NO \\
&&C_4=\frac{1}{\xi}A_T. \label{eqn:cbeta}
\eea

In collinear factorization, the protons interact with the virtual photons via the on-shell partons, and thus,
the hadronic tensor, $W_h^{\mu\nu}$, can be further factorized as the summation of the parton-level hadronic tensors
convoluted with the parton distribution functions (PDFs).
For $c\bar{c}(^3S_1^{[1]})$ production, there is only one parton-level process at QCD LO, \textit{i.e.}
\bea
g+\gamma^\star\rightarrow c\bar{c}(^3S_1^{[1]})+g, \label{eqn:processCS}
\eea
while for the production of the three color-octet states, namely $c\bar{c}(^1S_0^{[8]})$, $c\bar{c}(^3S_1^{[8]})$, and $c\bar{c}(^3P_J^{[8]})$,
there are two parton-level processes at QCD LO, they are
\bea
&&g+\gamma^\star\rightarrow c\bar{c}(n)+g, \NO \\
&&q~(\bar{q})+\gamma^\star\rightarrow c\bar{c}(n)+q~(\bar{q}). \label{eqn:processCO}
\eea
Here, $n$ represents $^1S_0^{[8]}$, $^3S_1^{[8]}$, or $^3P_J^{[8]}$.
Denoting the squared amplitudes of the above processes as $W_{i+\gamma^\star\rightarrow c\bar{c}(n)+i}^{\mu\nu}$,
where $\mu$ and $\nu$ are the Lorentz indices for the virtual photons in the amplitude and its complex conjugate, respectively,
$W_h^{\mu\nu}$ can be written as
\bea
W_h^{\mu\nu}=\sum_{i,n}\frac{1}{N_i}\int\frac{\md X}{X}f_{i/p}(X)W_{i+\gamma^\star\rightarrow c\bar{c}(n)+i}^{\mu\nu}
\langle\mathcal{O}^{J/\psi}(n)\rangle\md\Phi_X, \label{eqn:whdef}
\eea
where $n$ runs over the four intermediate $c\bar{c}$ states,
$N_i$ is the synthesized initial-state averaging factor,
$X$ is the fraction of the proton momentum taken by the parton,
$f_{i/p}$ is the corresponding PDF,
$\langle\mathcal{O}^{J/\psi}(n)\rangle$ is the LDME for an intermediate $c\bar{c}$ state, $c\bar{c}(n)$, producing a $J/\psi$,
and $\md\Phi_X$ is defined as
\bea
\md\Phi_X=(2\pi)^4\delta^4(q+p_i-p-p_i')\frac{\md^3p_i'}{(2\pi)^32p_{i0}'}. \label{eqn:phix}
\eea
Here, $p_i$ and $p_i'$ are the momenta of the initial- and final-state parton, respectively,
and $p_i=Xp$ has been implemented.
Having $\md X$ and $\md\Phi_X$ integrated over, $W_h^{\mu\nu}$ can be further simplified as (see e.g. Reference~\cite{Sun:2017nly})
\bea
W_h^{\mu\nu}=\frac{2\pi}{S}\sum_{i,n}\frac{f_{i/p}(X)}{X}\frac{1}{y(1-z)}
W_{i+\gamma^\star\rightarrow c\bar{c}(n)+i}^{\mu\nu}\langle\mathcal{O}^{J/\psi}(n)\rangle. \label{eqn:wh}
\eea

Having Equation~(\ref{eqn:wh}), Equation~(\ref{eqn:cs}) then leads to
\bea
\md\sigma=\frac{\alpha}{(4\pi)^3S}\sum_{i,n}\frac{f_{i/p}(X)}{X}l_{\mu\nu}W_{i+\gamma^\star\rightarrow c\bar{c}(n)+i}^{\mu\nu}
\langle\mathcal{O}^{J/\psi}(n)\rangle\frac{\md x}{x}\frac{\md y}{y}\frac{\md z}{z(1-z)}\md\xi\md\psi. \label{eqn:csr}
\eea
Note that in Equation~(\ref{eqn:csr}) the value of $X$ has been fixed by the 0-th dimension of the $\delta$-function at
\bea
X=x+\frac{\xi+\eta}{yz}+\frac{\xi}{y(1-z)}. \label{eqn:X}
\eea
To write the cross section in a more compact form, we define
\bea
&&W_1[n]=\alpha\sum_i\frac{f_{i/p}(X)}{X}(-g_{\mu\nu})W_{i+\gamma^\star\rightarrow c\bar{c}(n)+i}^{\mu\nu}, \NO \\
&&W_2[n]=\alpha\sum_i\frac{f_{i/p}(X)}{X}\frac{P_\mu P_\nu}{S}W_{i+\gamma^\star\rightarrow c\bar{c}(n)+i}^{\mu\nu}, \NO \\
&&W_3[n]=\alpha\sum_i\frac{f_{i/p}(X)}{X}\frac{P_\mu p_\nu+p_\mu P_\nu}{2S}W_{i+\gamma^\star\rightarrow c\bar{c}(n)+i}^{\mu\nu}, \NO \\
&&W_4[n]=\alpha\sum_i\frac{f_{i/p}(X)}{X}\frac{p_\mu p_\nu}{S}W_{i+\gamma^\star\rightarrow c\bar{c}(n)+i}^{\mu\nu}. \label{eqn:wn}
\eea
The cross section can correspondingly be expressed as
\bea
\md\sigma=\frac{1}{(4\pi)^3S}\sum_{n}\sum_{i=1}^4C_iW_i[n]\langle\mathcal{O}^{J/\psi}(n)\rangle
\frac{\md x}{x}\frac{\md y}{y}\frac{\md z}{z(1-z)}\md\xi\md\psi. \label{eqn:csf}
\eea

Since $W_{i+\gamma^\star\rightarrow c\bar{c}(n)+i}^{\mu\nu}$ can be easily evaluated via Feynman diagrams,
we have been equipped with all the elements needed in Equation~(\ref{eqn:csr}) for calculating the $J/\psi$ leptoproduction cross sections.

\subsection{The calculation of the azimuthal asymmetry modulations}

\begin{figure}
\includegraphics[scale=0.35]{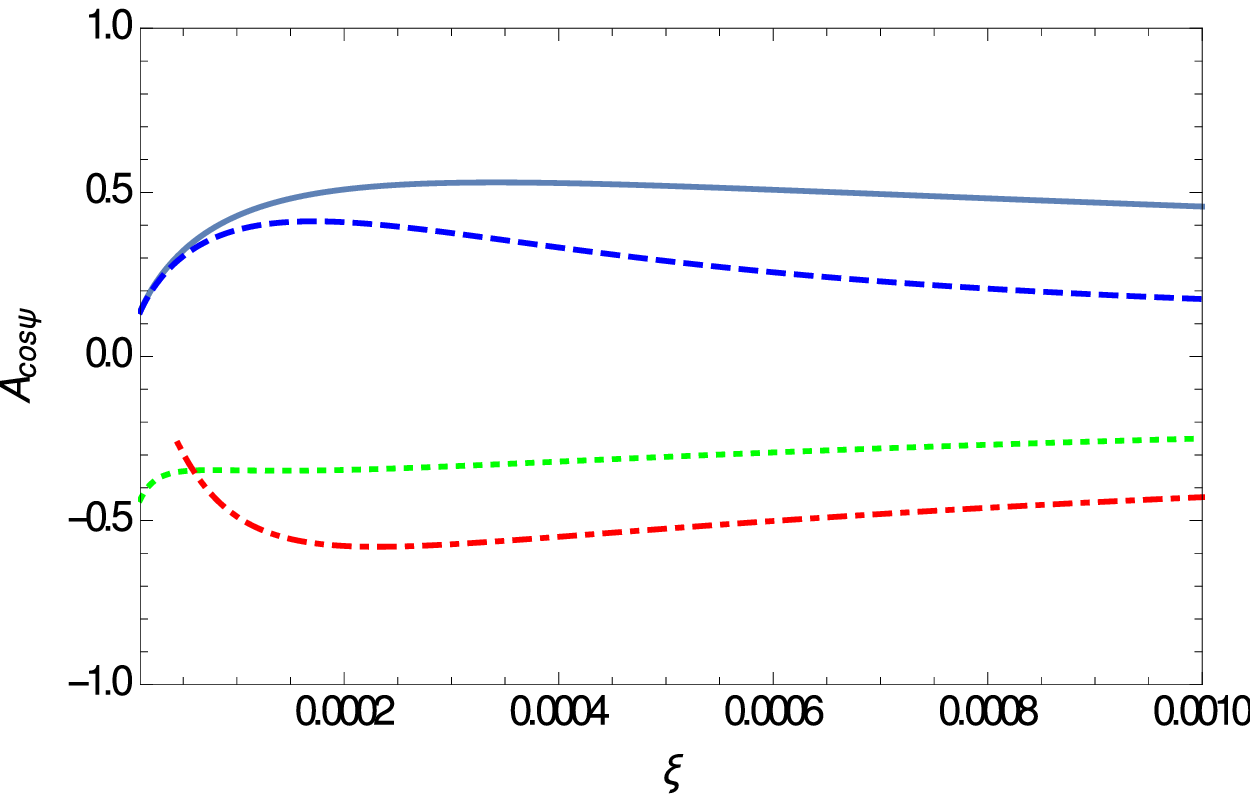}
\includegraphics[scale=0.35]{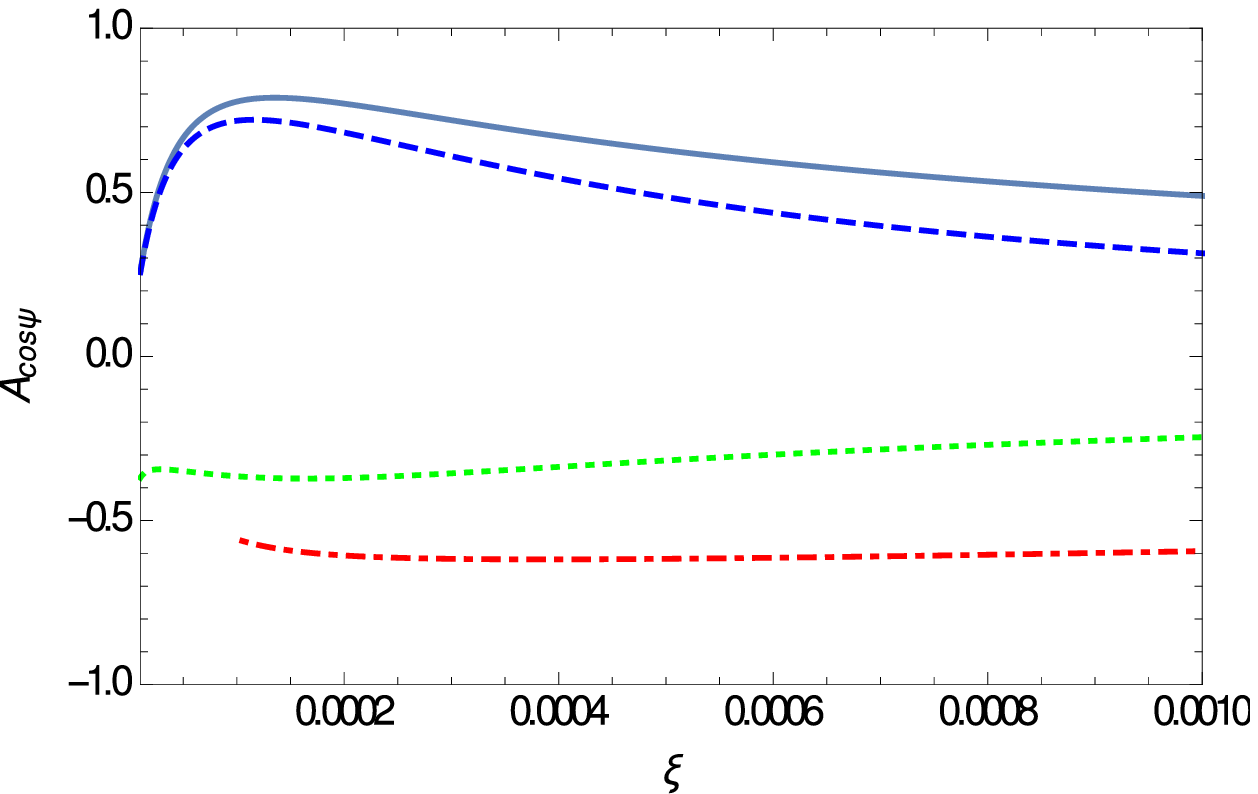}
\includegraphics[scale=0.35]{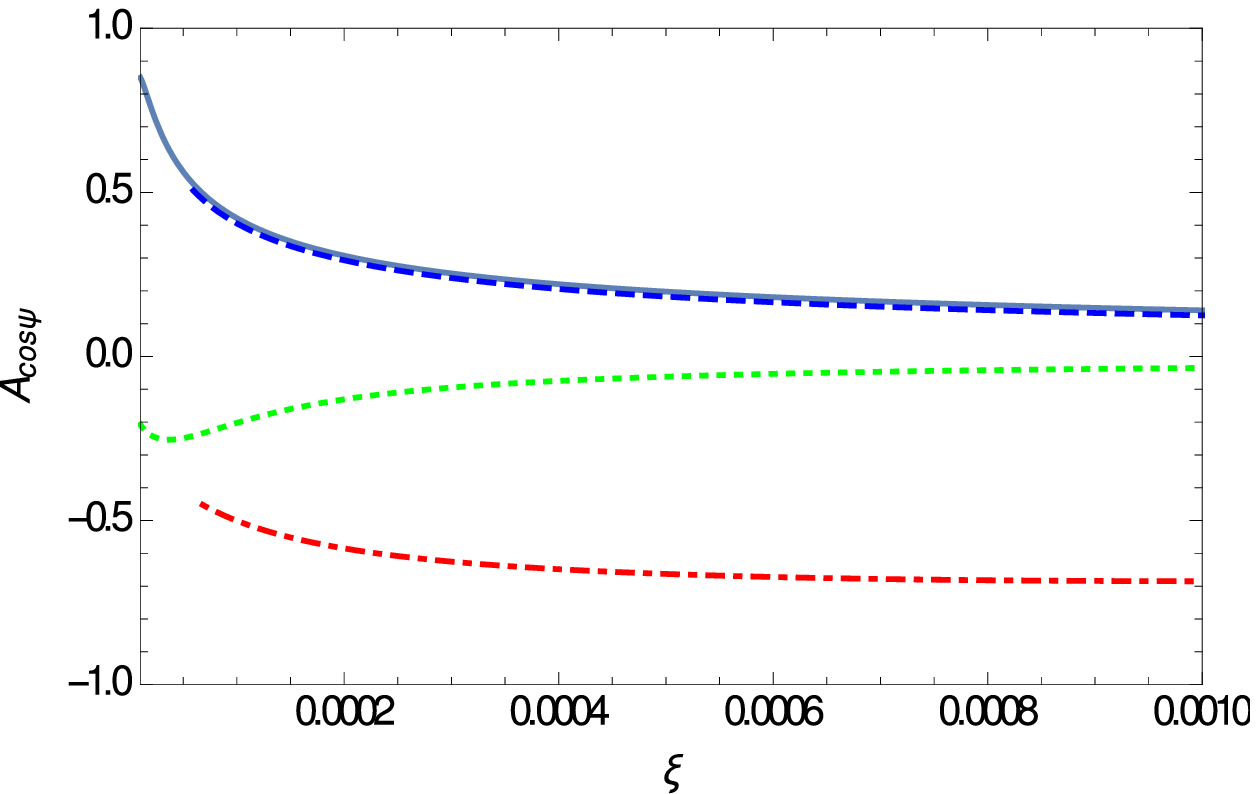}\\
\includegraphics[scale=0.35]{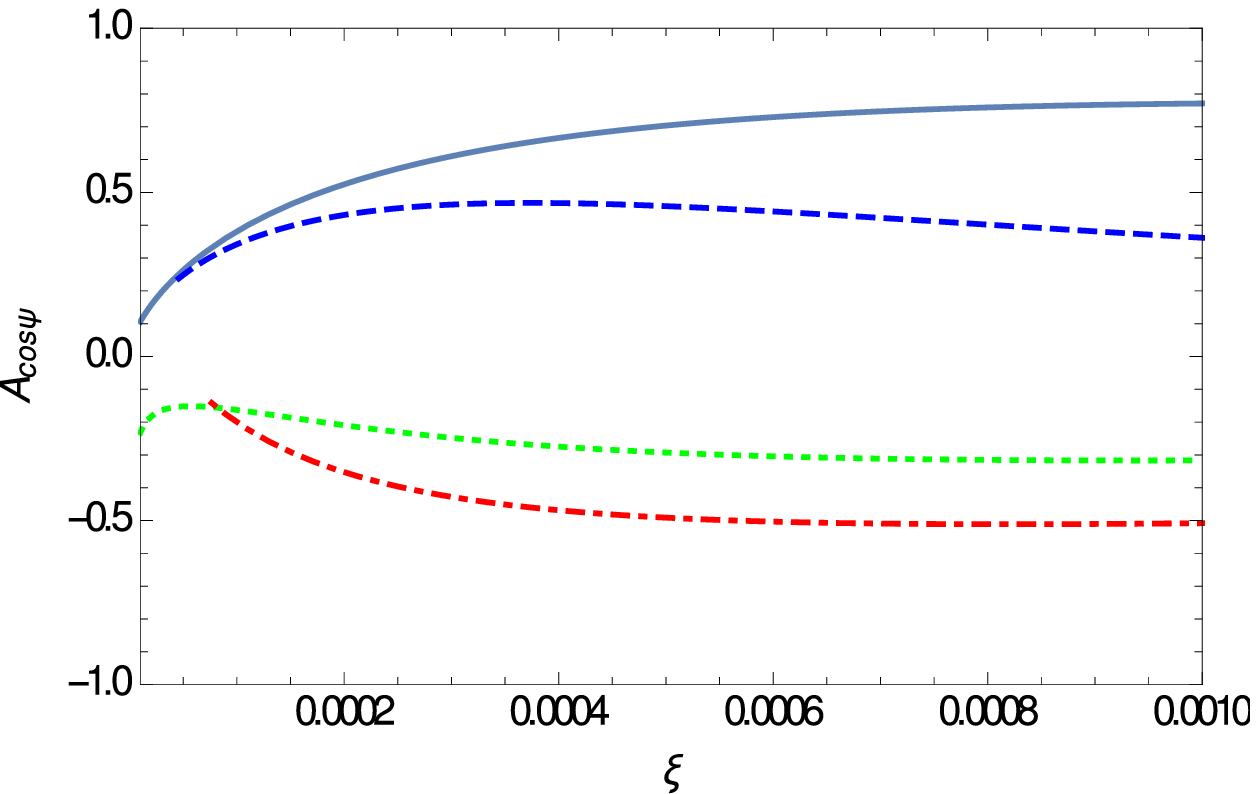}
\includegraphics[scale=0.35]{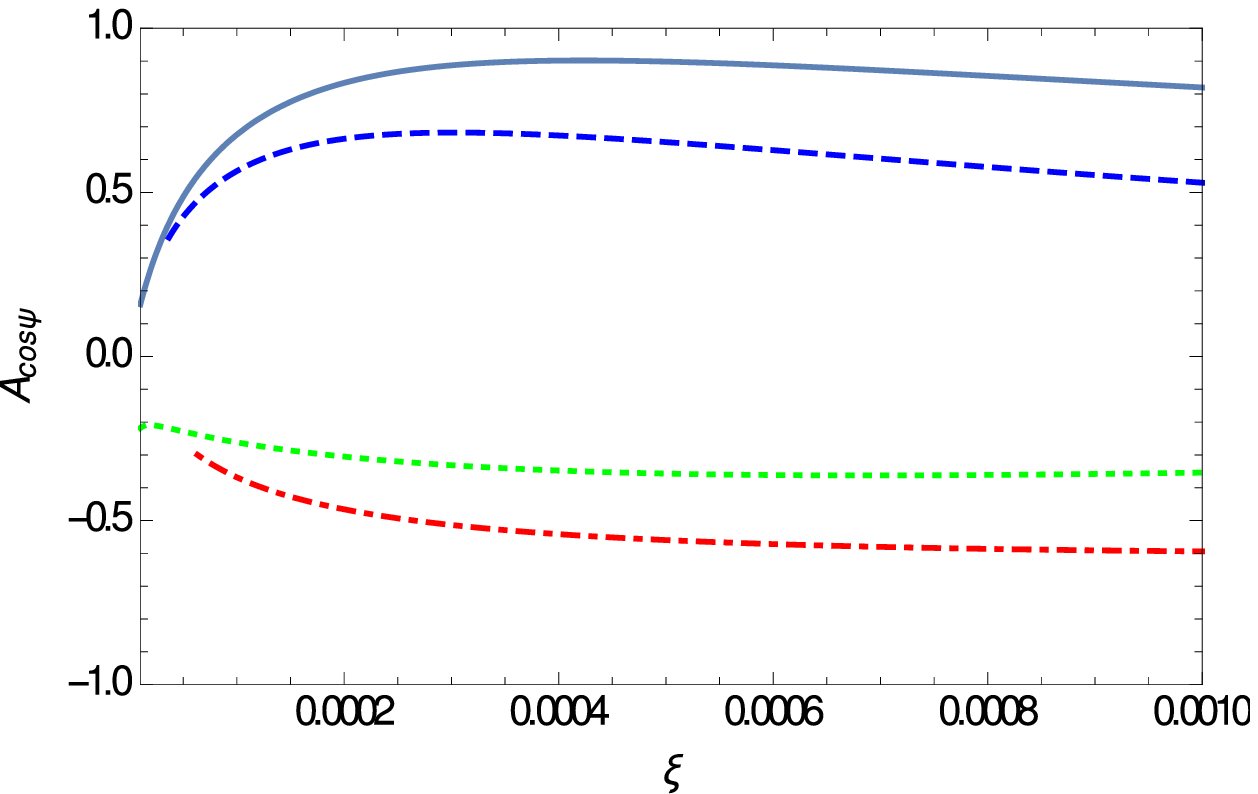}
\includegraphics[scale=0.35]{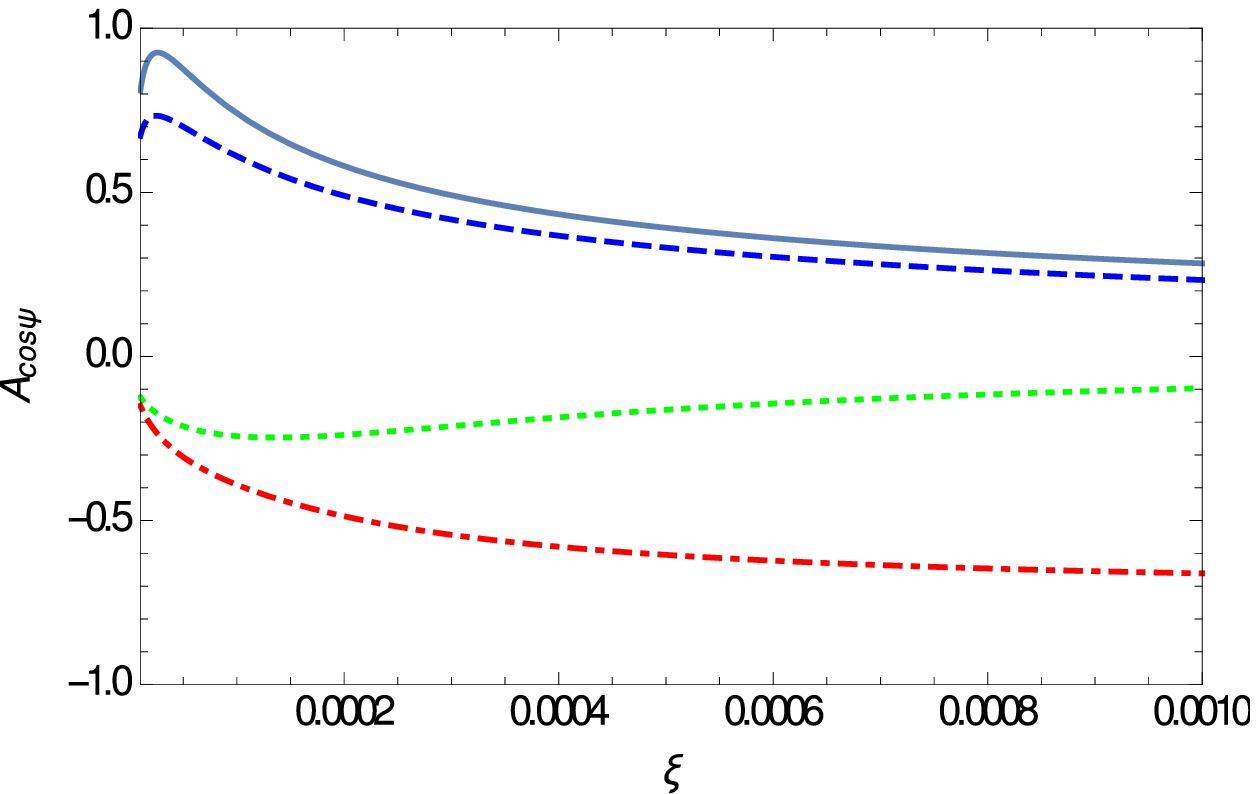}\\
\includegraphics[scale=0.35]{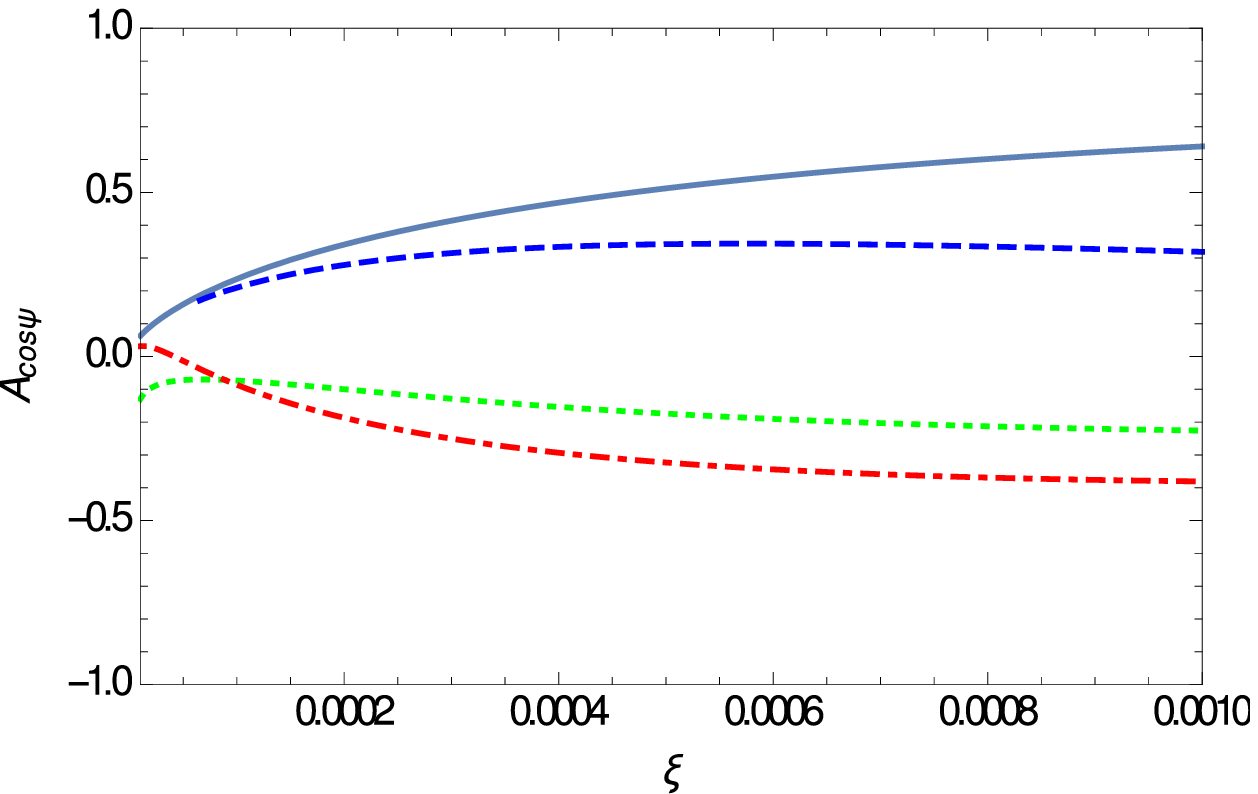}
\includegraphics[scale=0.35]{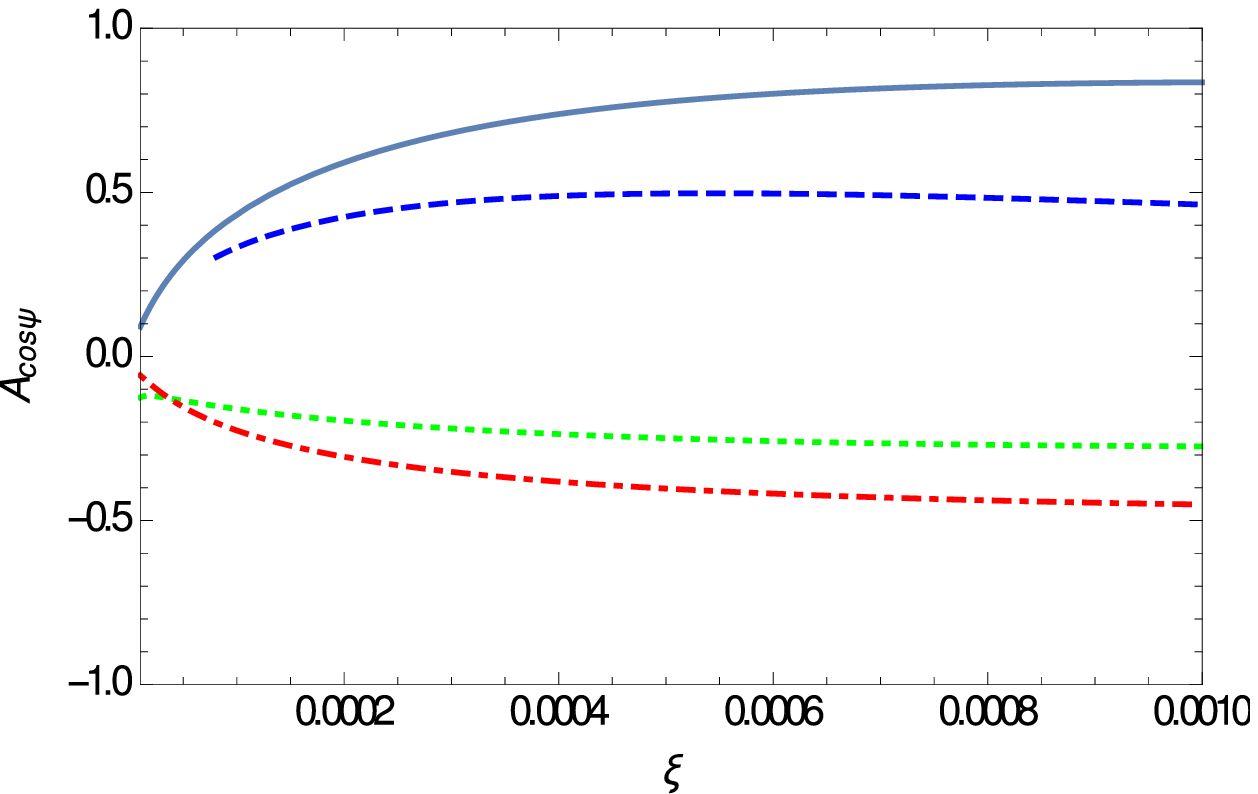}
\includegraphics[scale=0.35]{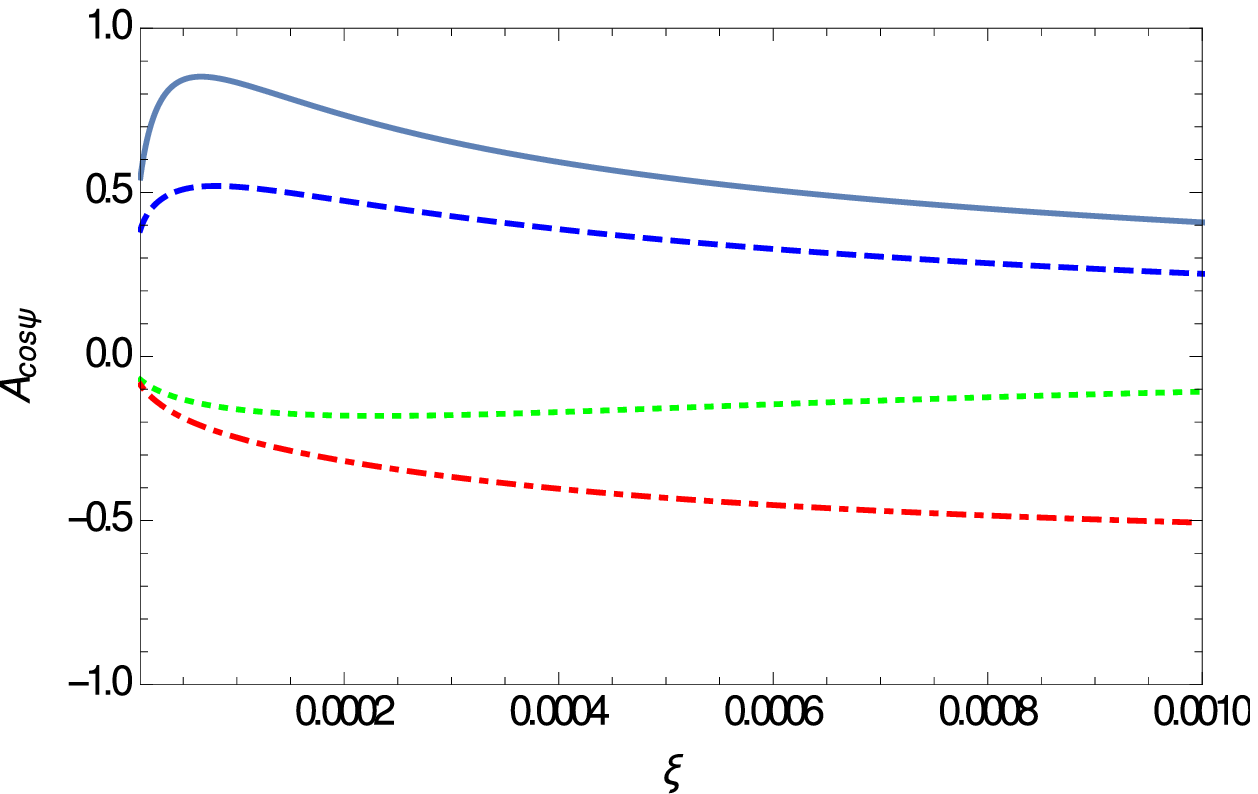}
\caption{\label{fig:cm11}
The $\mathrm{cos}(\psi)$ modulation as functions of $\xi$ for $x=0.005$.
The upper, mid, and lower rows correspond to $y=0.1$, $0.4$, and $0.7$, respectively,
while the left, mid, and right columns correspond to $z=0.3$, $0.6$, and $0.9$, respectively.
The solid, dotted, dashed, dashdotted curves correspond to the results for the $^3S_1^{[1]}$,
$^1S_0^{[8]}$, $^3S_1^{[8]}$, and $^3P_J^{[8]}$ states, respectively.
}
\end{figure}

Since $W_{i+\gamma^\star\rightarrow c\bar{c}(n)+i}^{\mu\nu}$ and the tensor basis in Equation~(\ref{eqn:cluv}) are all restricted in the hadronic plane,
\textit{i.e.} they are independent of the azimuthal angle $\psi$,
the azimuthal asymmetry modulations appear only in the coefficients in the leptonic tensor expansions, namely in $C_i$'s.
In any case, the cross section can be written in the following form,
\bea
\md\sigma=\sigma_0\left[1+A_{\mathrm{cos}\psi}\mathrm{cos}(\psi)+A_{\mathrm{cos}2\psi}\mathrm{cos}(2\psi)\right]
\md x\md y\md z\md\xi\md\psi. \label{eqn:csm}
\eea
Therein, $A_{\mathrm{cos}\psi}$ and $A_{\mathrm{cos}2\psi}$ are two independent azimuthal asymmetry modulations.
They are functions of $x$, $y$, $z$, and $\xi$.
Explicitly, they can be accessed via the following equations,
\bea
&&A_{\mathrm{cos}\psi}=\frac{2\int_0^{2\pi}\md\psi\mathrm{cos}(\psi)\frac{\md\sigma}{\md x\md y\md z\md\xi}}
{\int_0^{2\pi}\md\psi\frac{\md\sigma}{\md x\md y\md z\md\xi}}, \NO \\
&&A_{\mathrm{cos}2\psi}=\frac{2\int_0^{2\pi}\md\psi\mathrm{cos}(2\psi)\frac{\md\sigma}{\md x\md y\md z\md\xi}}
{\int_0^{2\pi}\md\psi\frac{\md\sigma}{\md x\md y\md z\md\xi}}. \label{eqn:adef}
\eea

Since the four intermediate $c\bar{c}$ states contribute independently,
one can study the azimuthal asymmetry originated from them respectively.
In this sense, the cross section can be written in the following form as
\bea
\md\sigma=\sum_n\sigma_0^n\left[1+A_{\mathrm{cos}\psi}^n\mathrm{cos}(\psi)+A_{\mathrm{cos}2\psi}^n\mathrm{cos}(2\psi)\right]
\md x\md y\md z\md\xi\md\psi. \label{eqn:csmn}
\eea
In the next section as we present their numerical results,
we just omit the superscript $n$.

\begin{figure}
\includegraphics[scale=0.35]{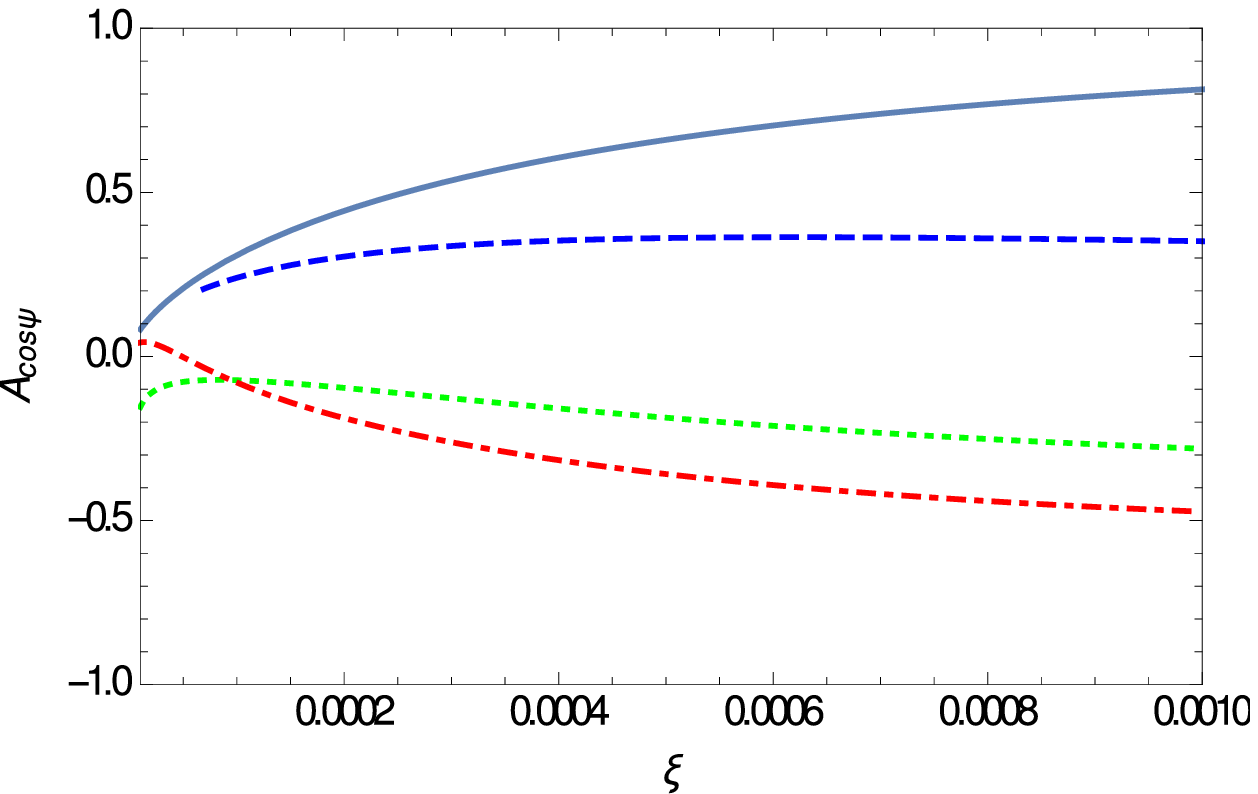}
\includegraphics[scale=0.35]{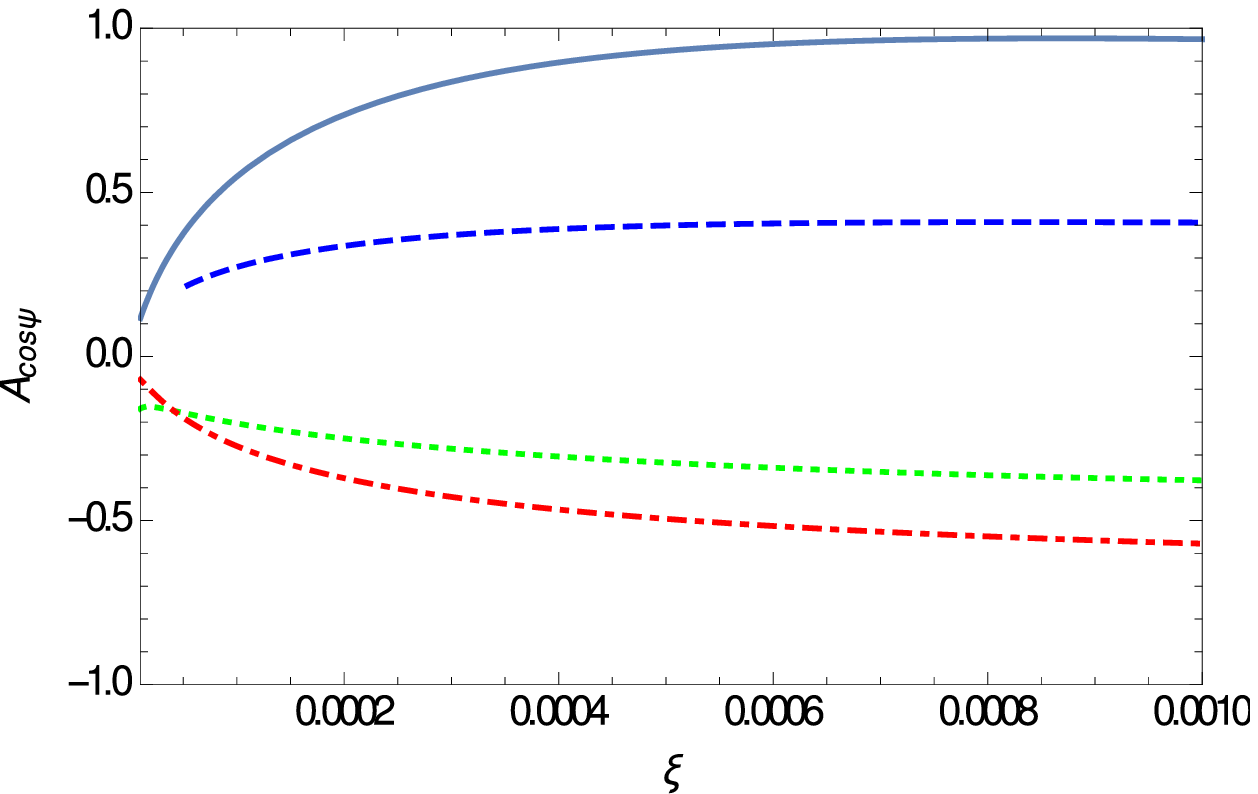}
\includegraphics[scale=0.35]{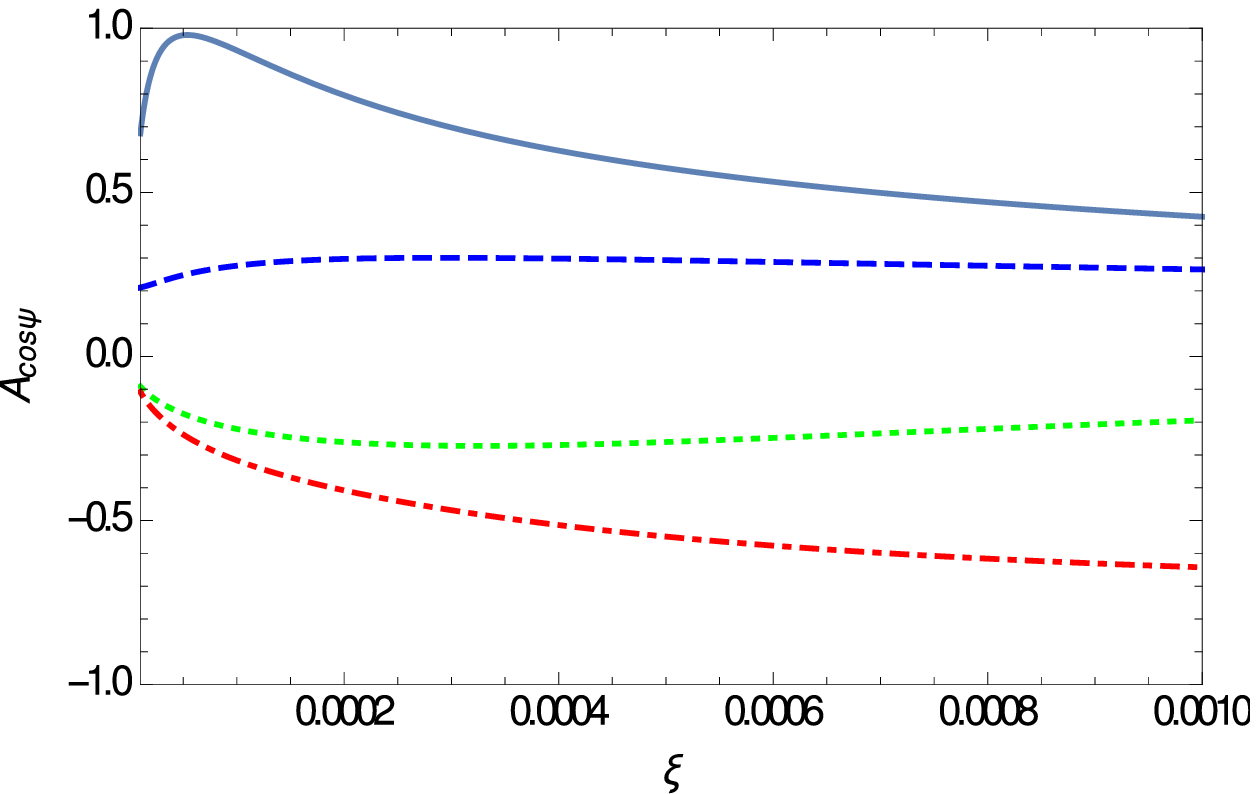}\\
\includegraphics[scale=0.35]{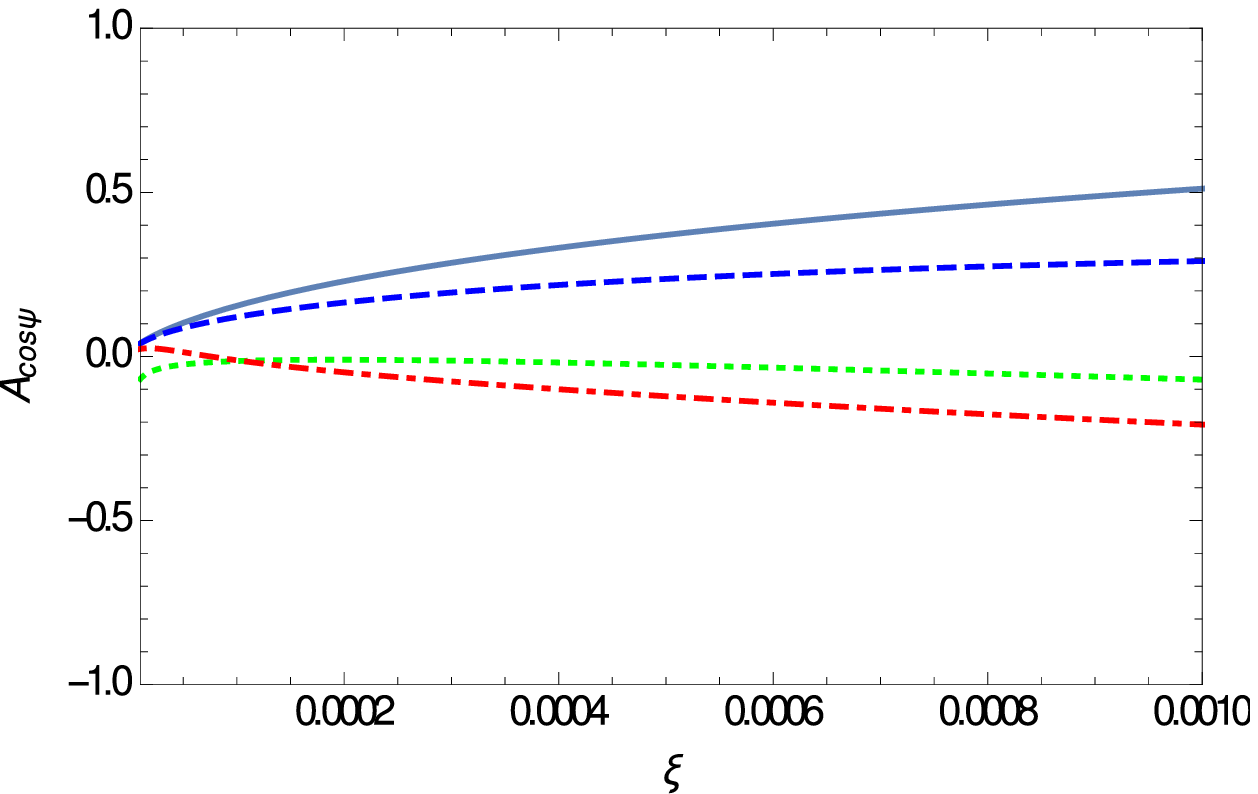}
\includegraphics[scale=0.35]{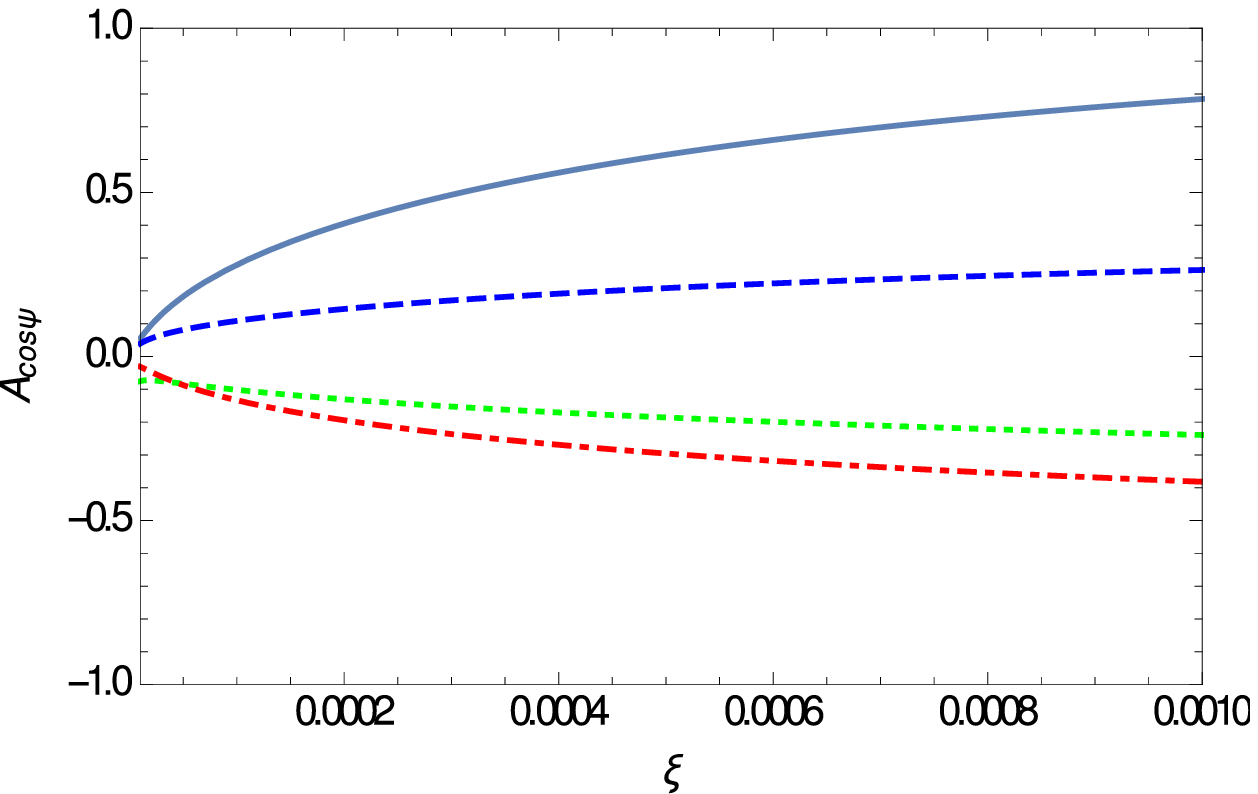}
\includegraphics[scale=0.35]{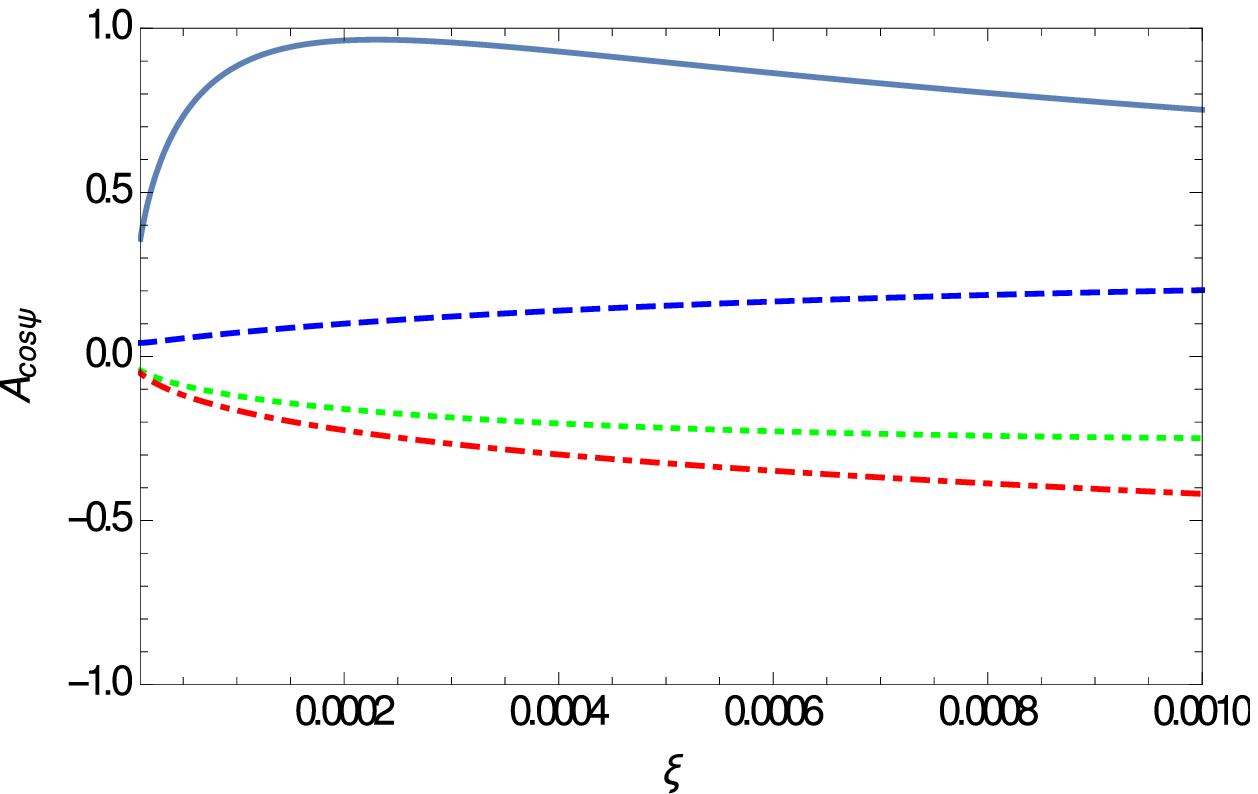}\\
\includegraphics[scale=0.35]{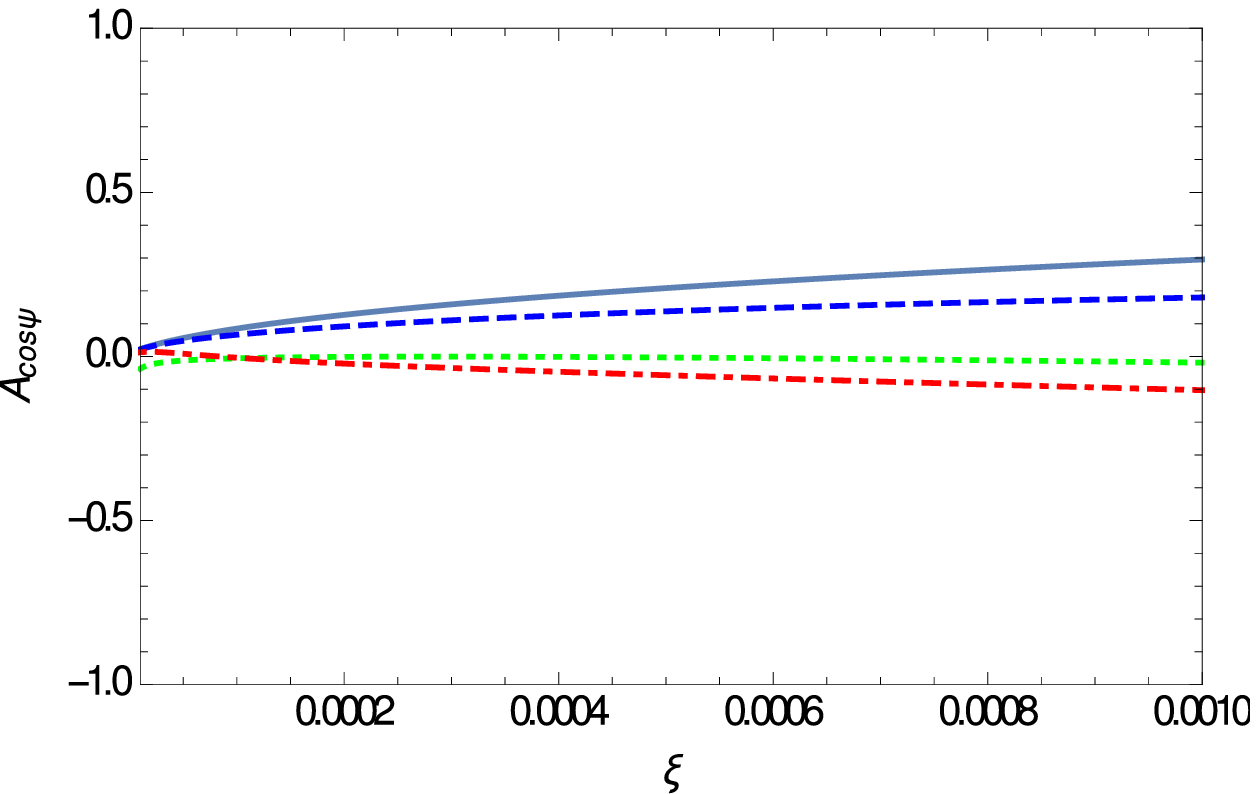}
\includegraphics[scale=0.35]{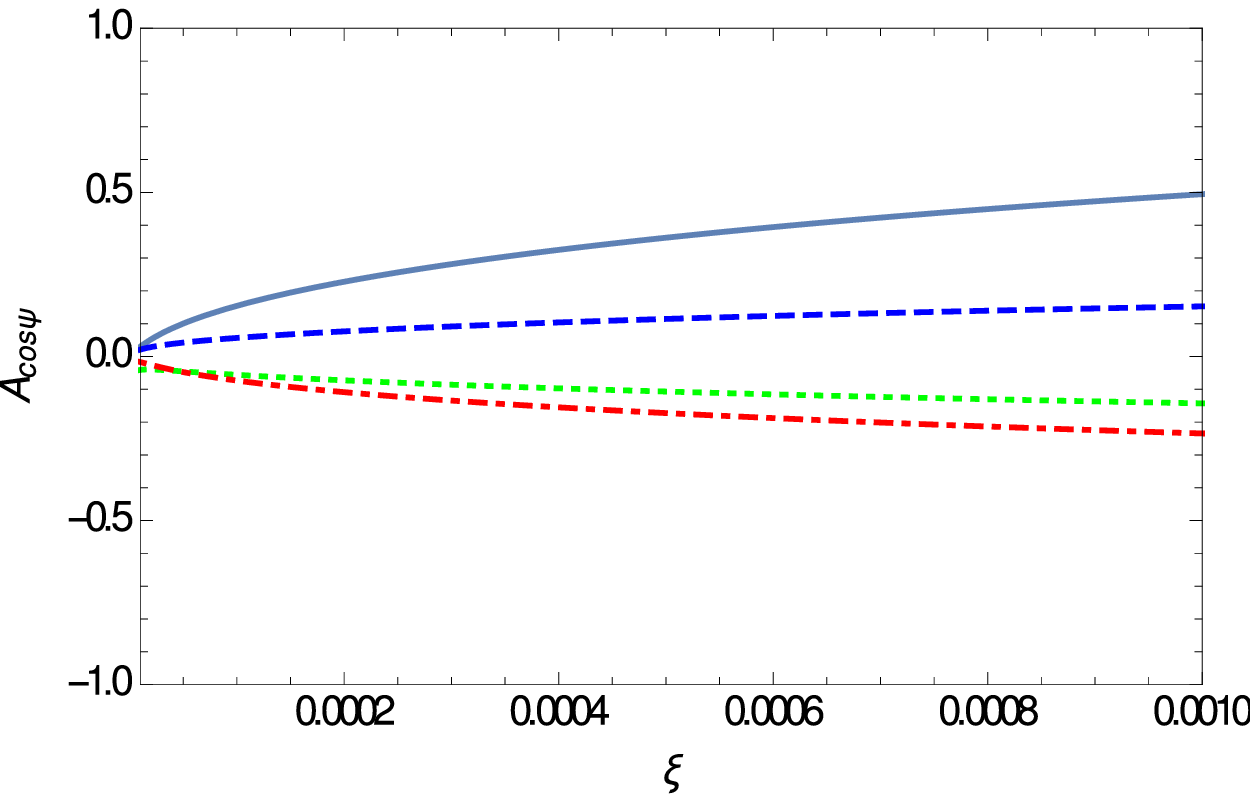}
\includegraphics[scale=0.35]{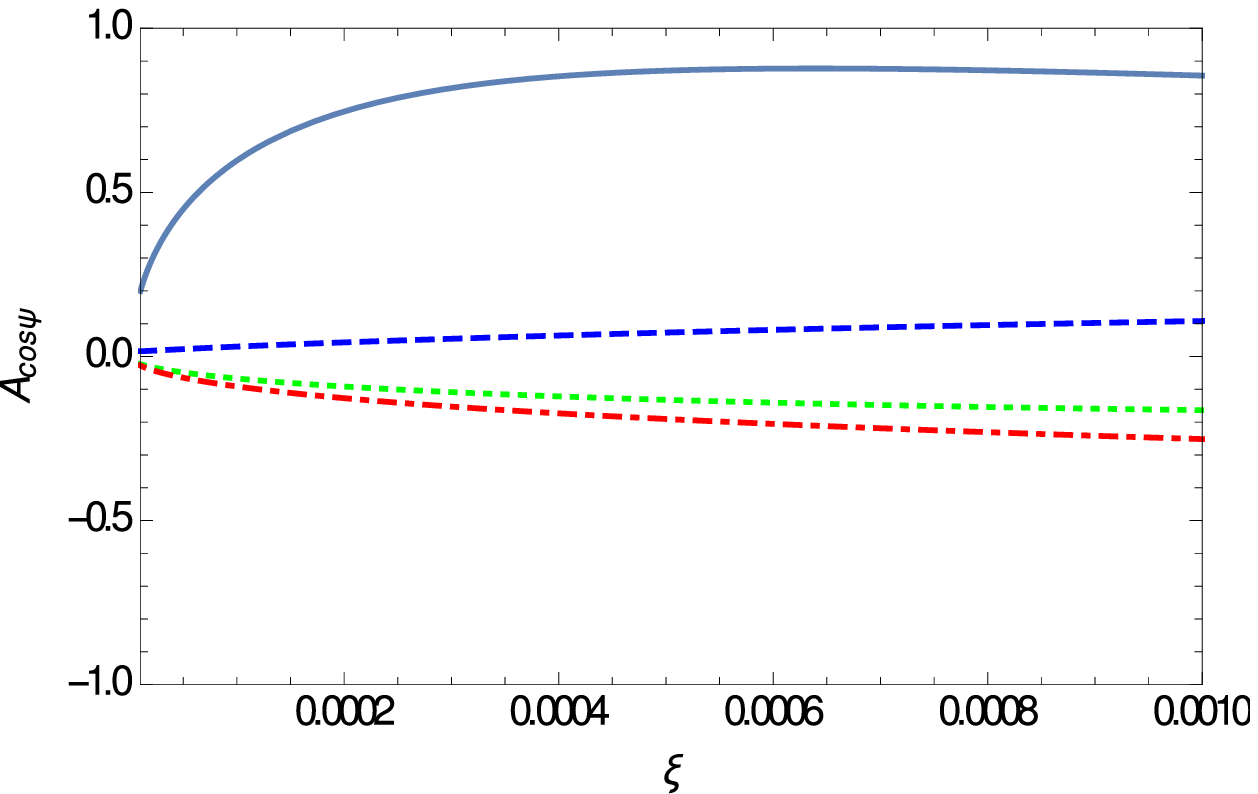}
\caption{\label{fig:cm12}
The $\mathrm{cos}(\psi)$ modulation as functions of $\xi$ for $x=0.05$.
The upper, mid, and lower rows correspond to $y=0.1$, $0.4$, and $0.7$, respectively,
while the left, mid, and right columns correspond to $z=0.3$, $0.6$, and $0.9$, respectively.
The solid, dotted, dashed, dashdotted curves correspond to the results for the $^3S_1^{[1]}$,
$^1S_0^{[8]}$, $^3S_1^{[8]}$, and $^3P_J^{[8]}$ states, respectively.
}
\end{figure}

Sometimes, it is useful to study the integrated cross sections,
the azimuthal asymmetry modulations should be redefined accordingly.
When we observe the azimuthal asymmetry modulations in a specific kinematic region, \textit{e.g} $D$,
where $D$ is a 4-dimensional area in the $x$-$y$-$z$-$\xi$ space,
the corresponding azimuthal asymmetry modulations in the region $D$, $A_{\mathrm{cos}\psi}^D$ and $A_{\mathrm{cos}2\psi}^D$,
are defined according to the following equation,
\bea
\md\sigma^D=\sigma_0^D\left[1+A_{\mathrm{cos}\psi}^D\mathrm{cos}(\psi)+A_{\mathrm{cos}2\psi}^D\mathrm{cos}(2\psi)\right]\md\psi. \label{eqn:aac}
\eea
They can also be expressed in the following explicit form as
\bea
&&A_{\mathrm{cos}\psi}^D=\frac{2\int_0^{2\pi}\md\psi\mathrm{cos}(\psi)\int_D\md x\md y\md z\md\xi\frac{\md\sigma}{\md x\md y\md z\md\xi\md\psi}}
{\int_0^{2\pi}\md\psi\int_D\md x\md y\md z\md\xi\frac{\md\sigma}{\md x\md y\md z\md\xi\md\psi}}, \NO \\
&&A_{\mathrm{cos}2\psi}^D=\frac{2\int_0^{2\pi}\md\psi\mathrm{cos}(2\psi)\int_D\md x\md y\md z\md\xi\frac{\md\sigma}{\md x\md y\md z\md\xi\md\psi}}
{\int_0^{2\pi}\md\psi\int_D\md x\md y\md z\md\xi\frac{\md\sigma}{\md x\md y\md z\md\xi\md\psi}}. \label{eqn:aai}
\eea

\section{Numerical results and phenomenological analysis\label{sec:numerical}}

\begin{figure}
\includegraphics[scale=0.35]{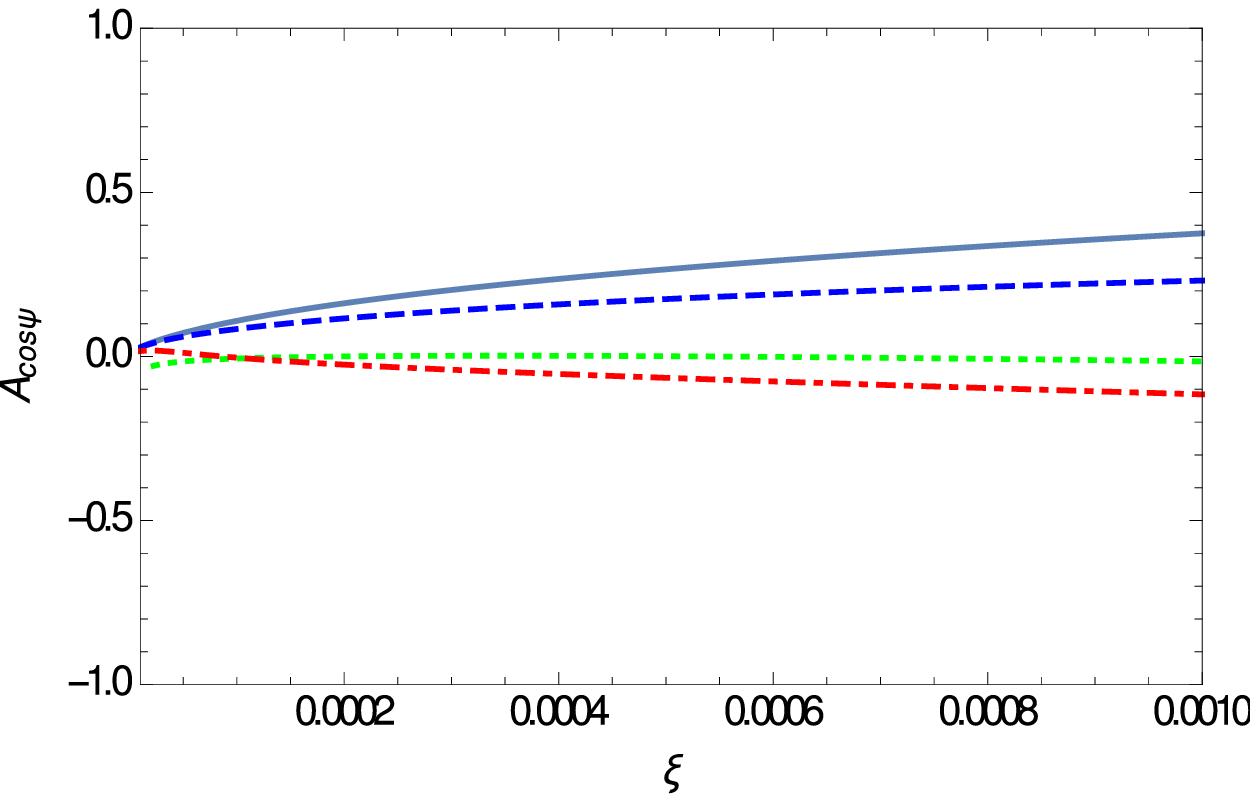}
\includegraphics[scale=0.35]{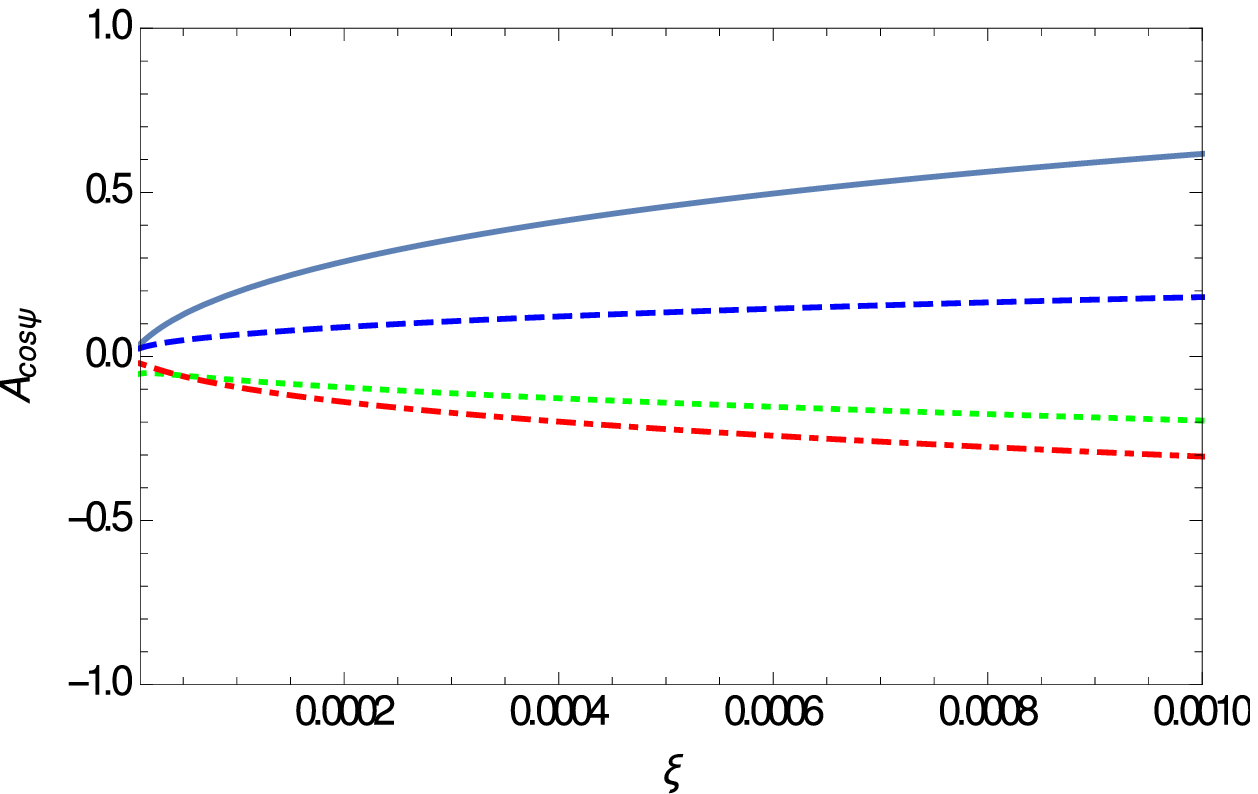}
\includegraphics[scale=0.35]{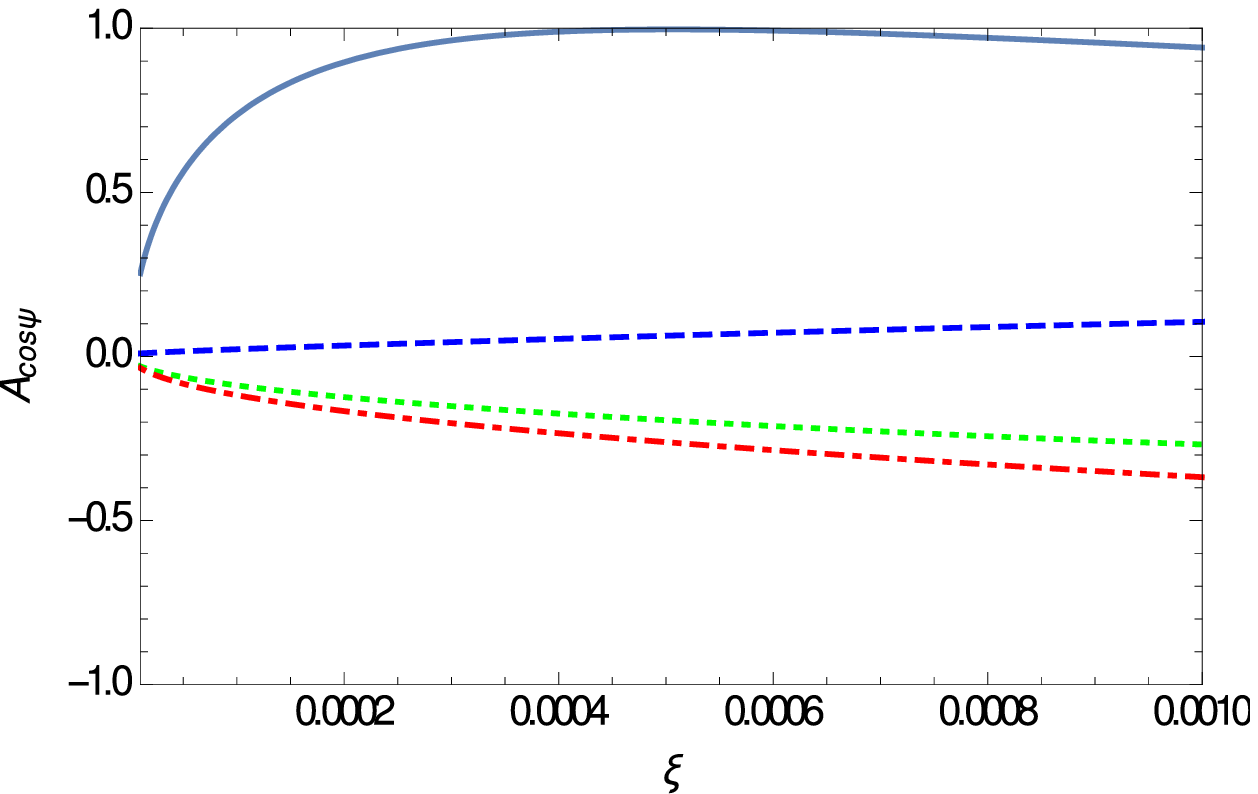}\\
\includegraphics[scale=0.35]{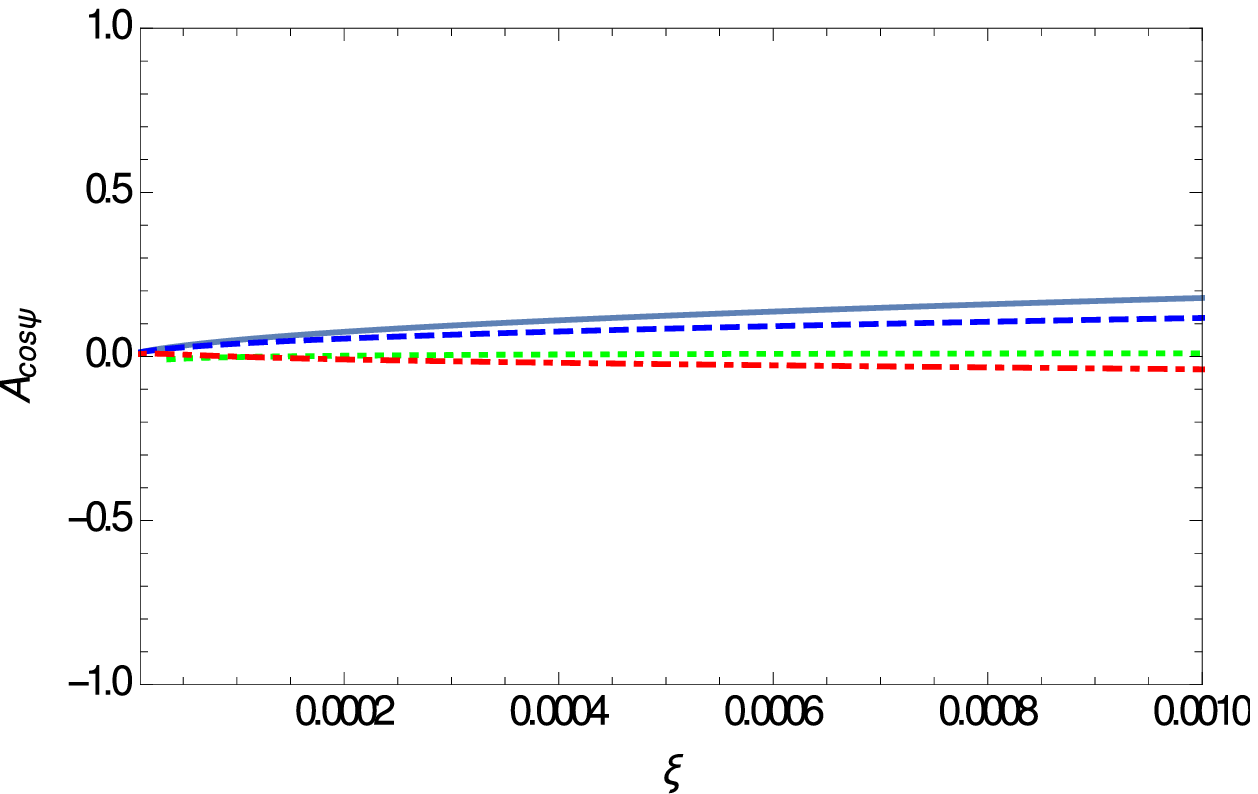}
\includegraphics[scale=0.35]{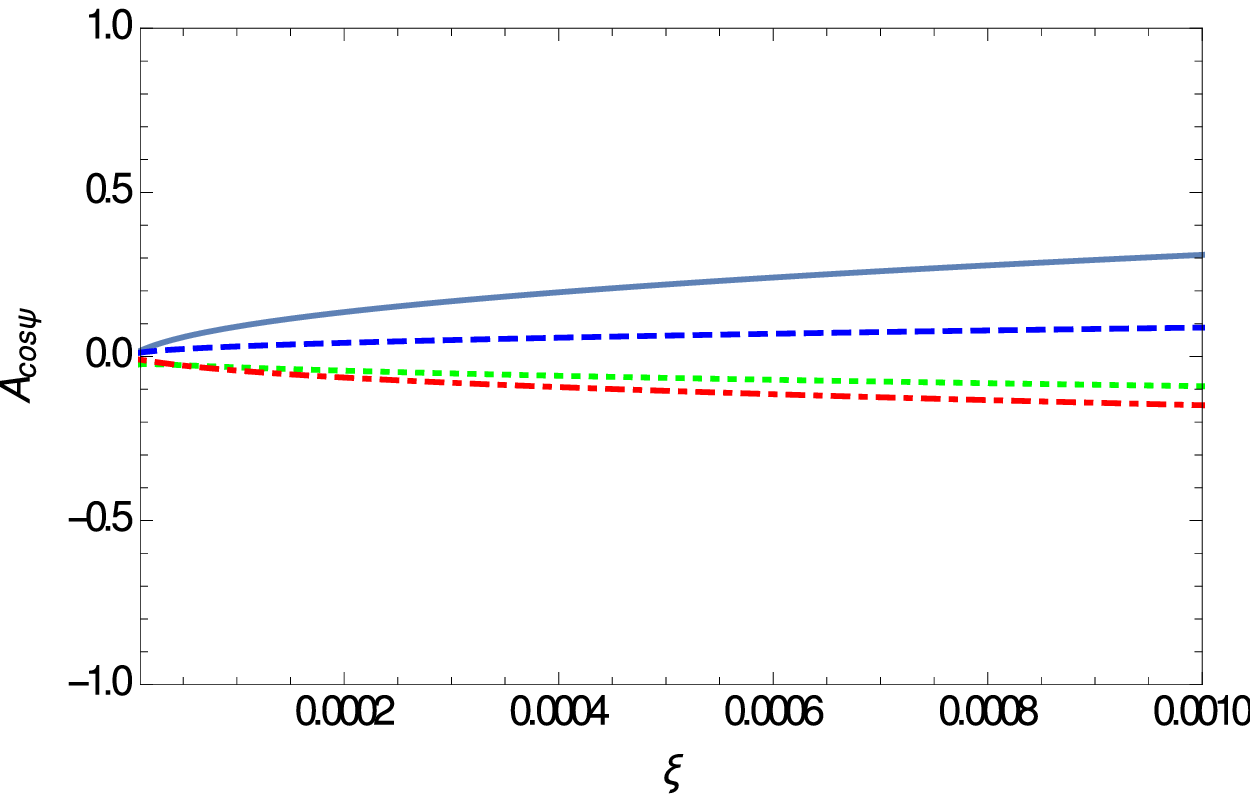}
\includegraphics[scale=0.35]{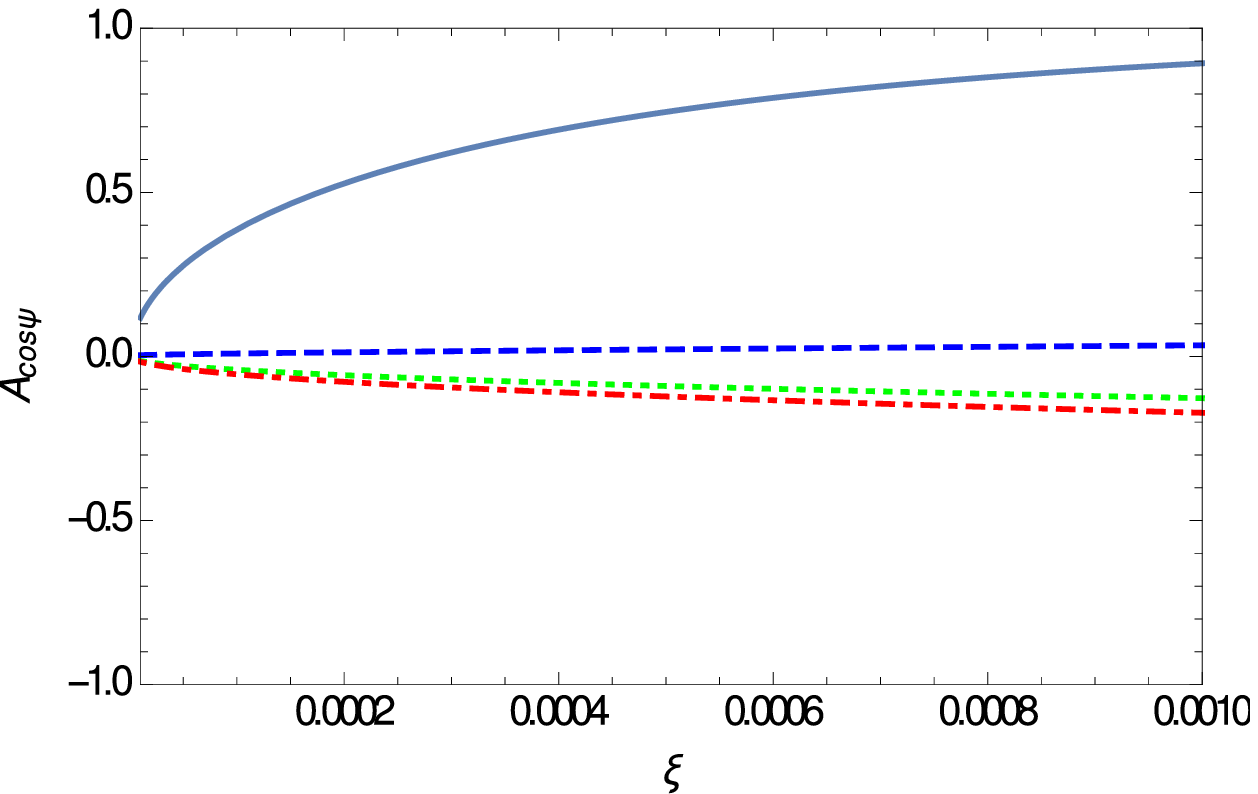}\\
\includegraphics[scale=0.35]{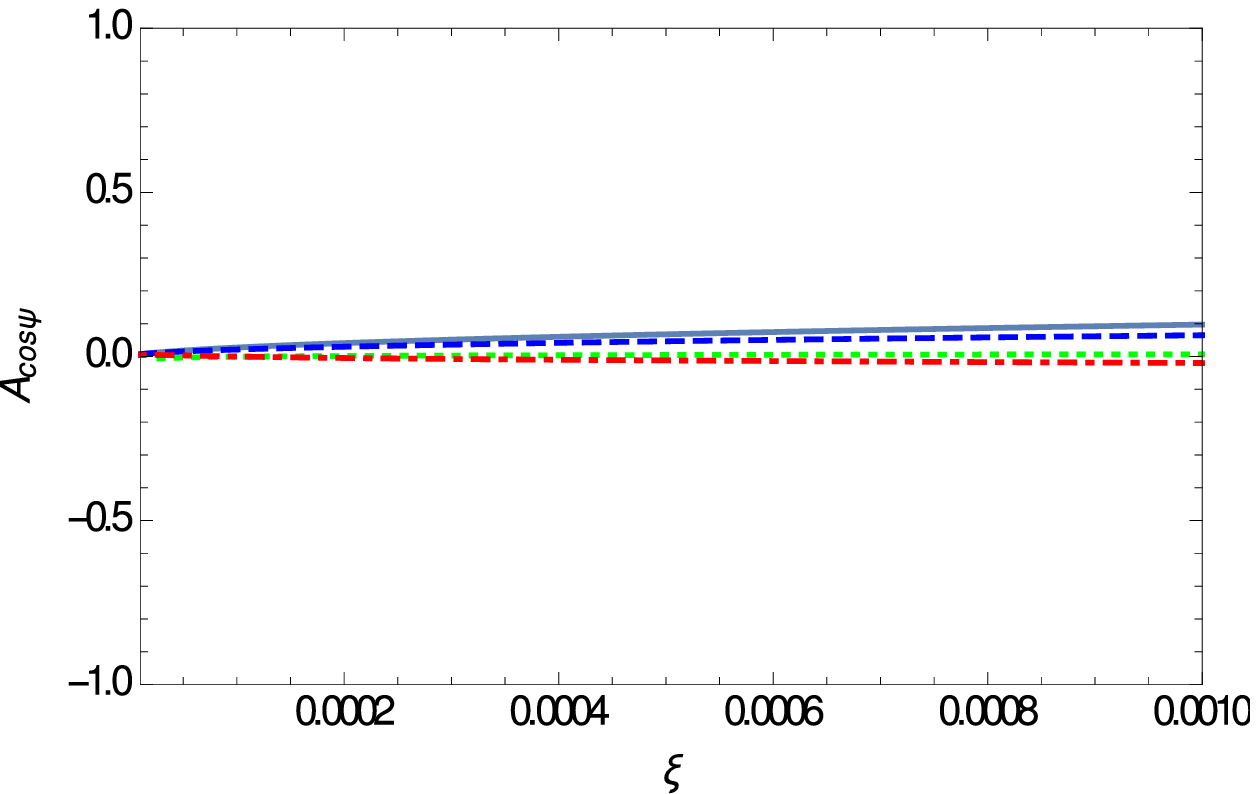}
\includegraphics[scale=0.35]{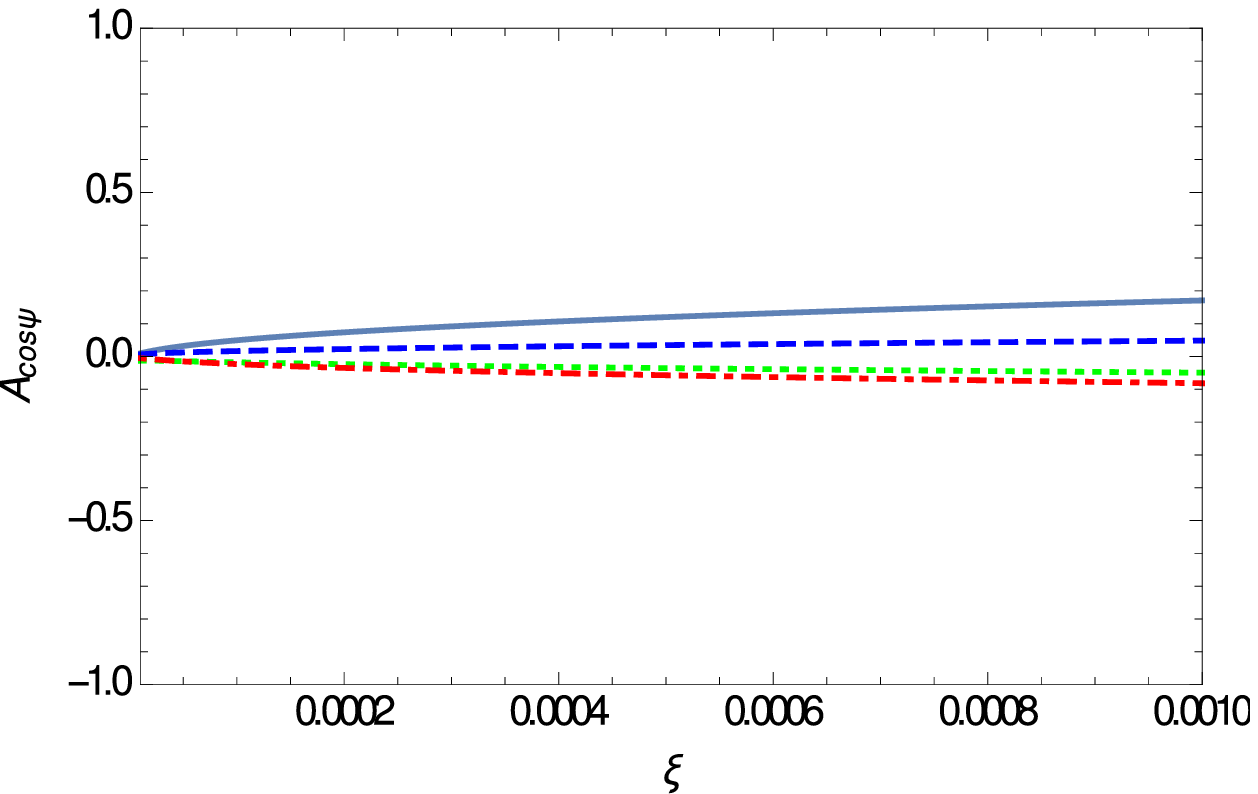}
\includegraphics[scale=0.35]{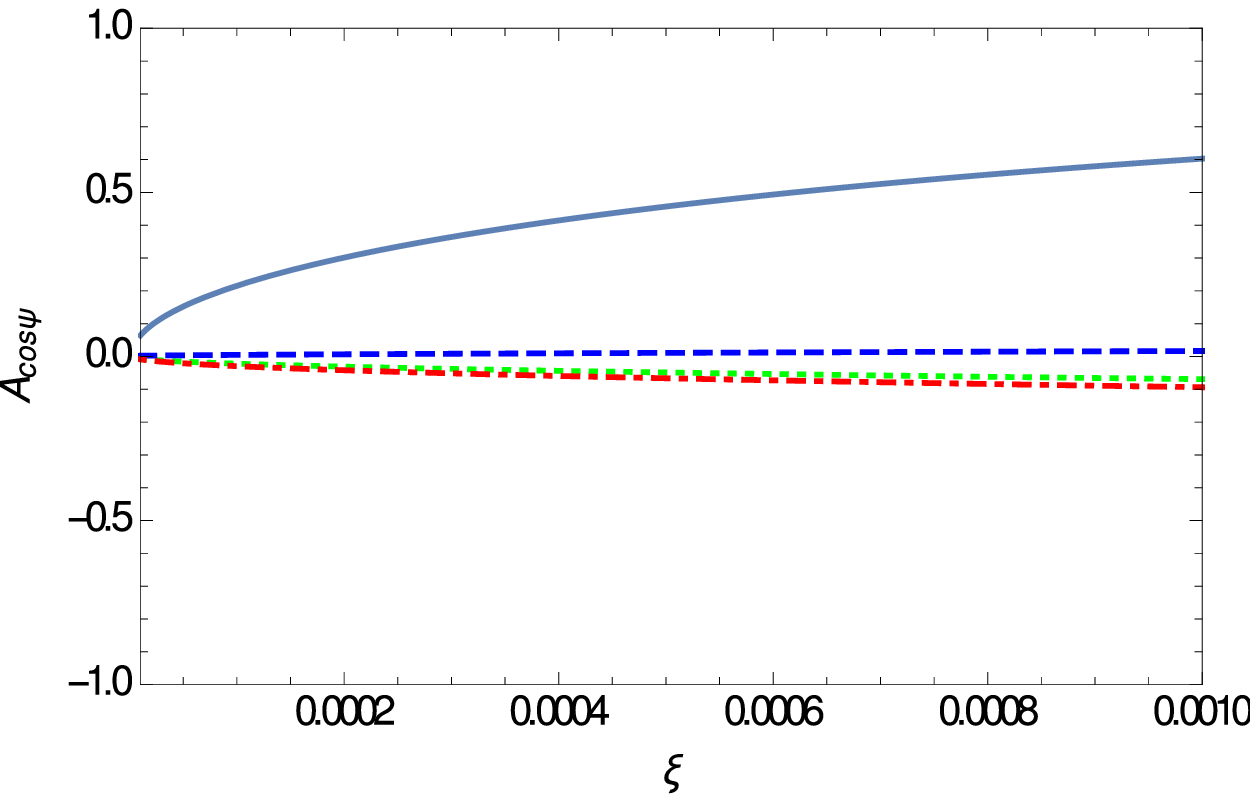}
\caption{\label{fig:cm13}
The $\mathrm{cos}(\psi)$ modulation as functions of $\xi$ for $x=0.5$.
The upper, mid, and lower rows correspond to $y=0.1$, $0.4$, and $0.7$, respectively,
while the left, mid, and right columns correspond to $z=0.3$, $0.6$, and $0.9$, respectively.
The solid, dotted, dashed, dashdotted curves correspond to the results for the $^3S_1^{[1]}$,
$^1S_0^{[8]}$, $^3S_1^{[8]}$, and $^3P_J^{[8]}$ states, respectively.
}
\end{figure}

In our numerical calculation, we dopt the following parameter choices.
The charm quark mass, $m_c$, is fixed at one half of the $J/\psi$ mass,
which is approximated to $M=3.0\gev$.
The colliding energy is chosen as $\sqrt{S}=318\gev$ to accord with the HERA experiment.
To evaluate the parton distributions in protons, we employ the GRV PDF given in Reference~\cite{Gluck:1994uf}.
Therein, the factorization scale is set to be $\mu_f=\sqrt{Q^2+p_t^2}=\sqrt{S(xy+\xi)}$.

\begin{figure}
\includegraphics[scale=0.35]{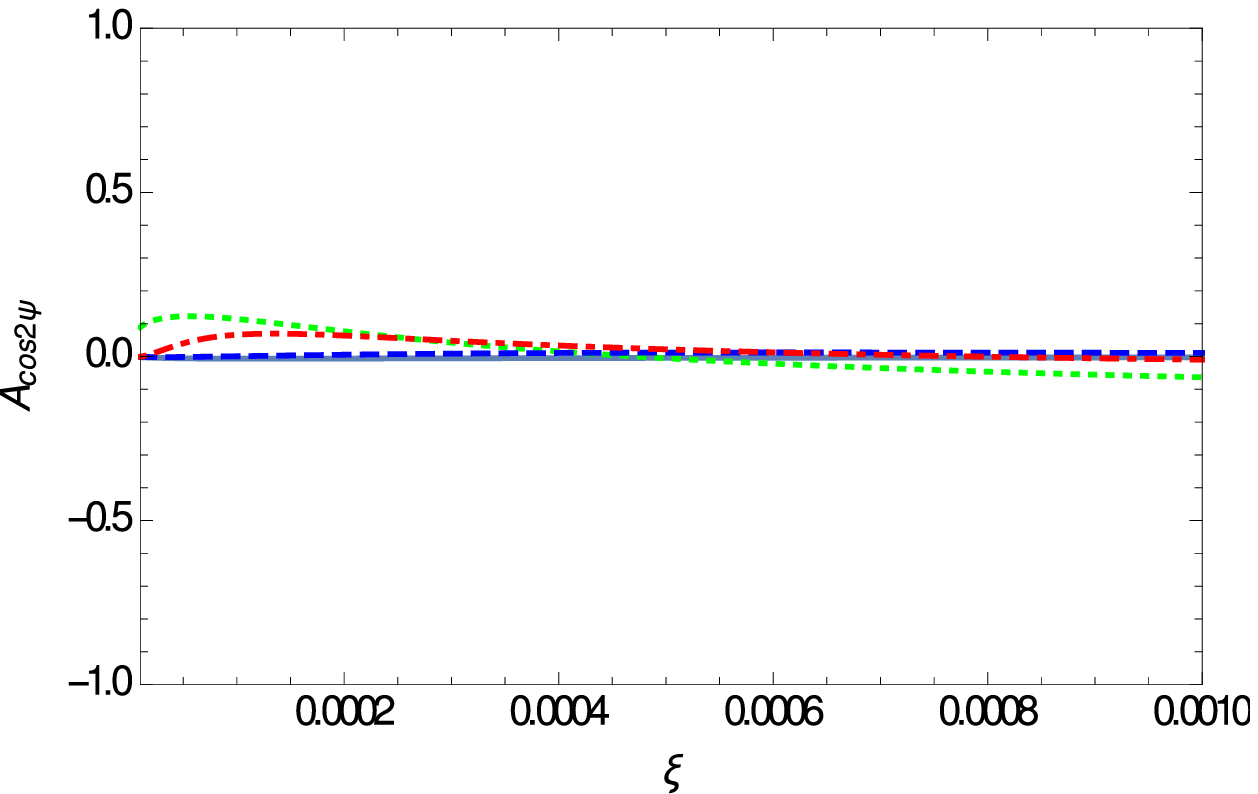}
\includegraphics[scale=0.35]{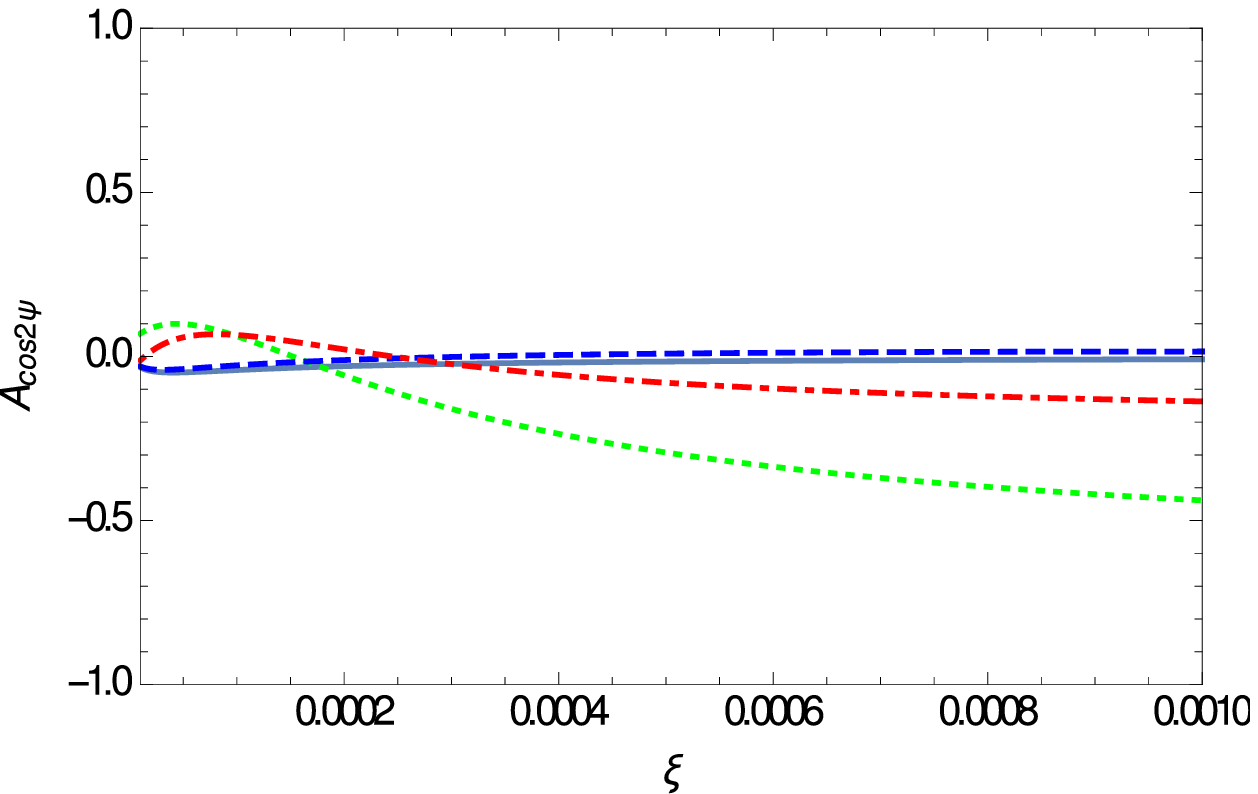}
\includegraphics[scale=0.35]{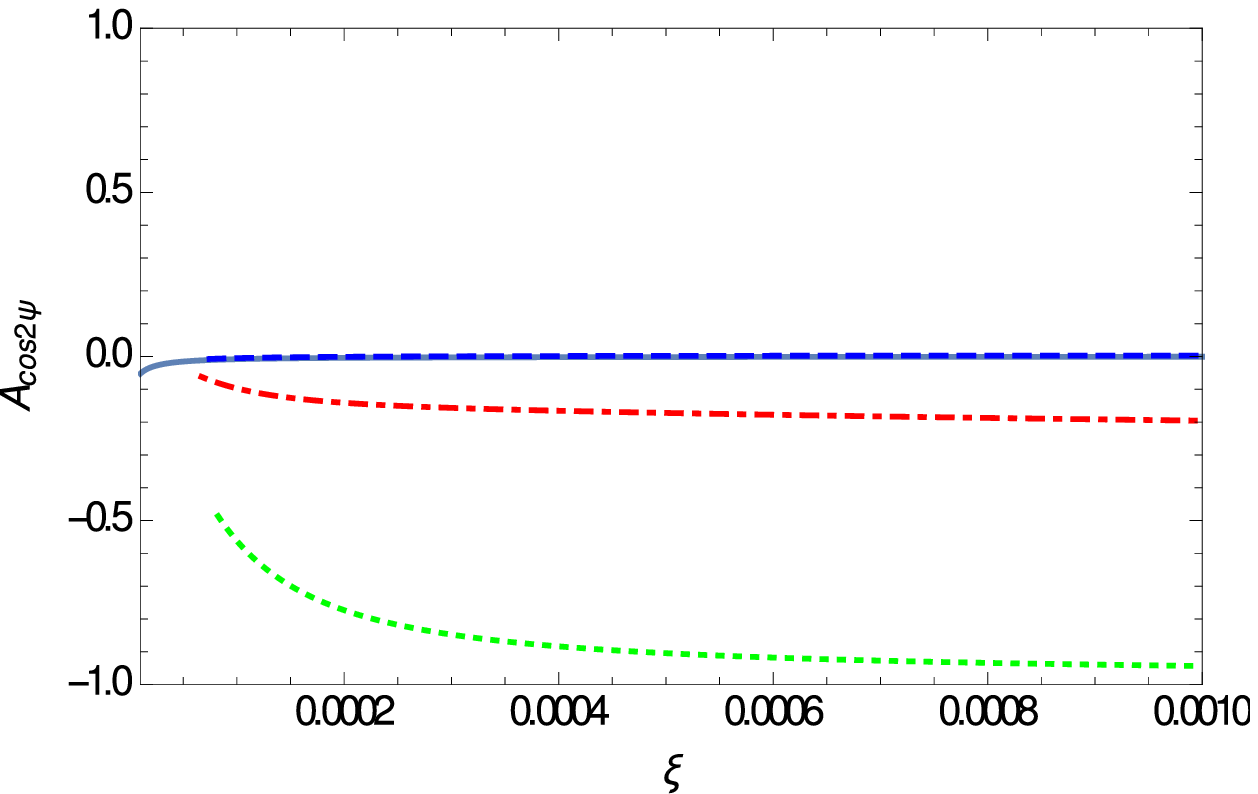}\\
\includegraphics[scale=0.35]{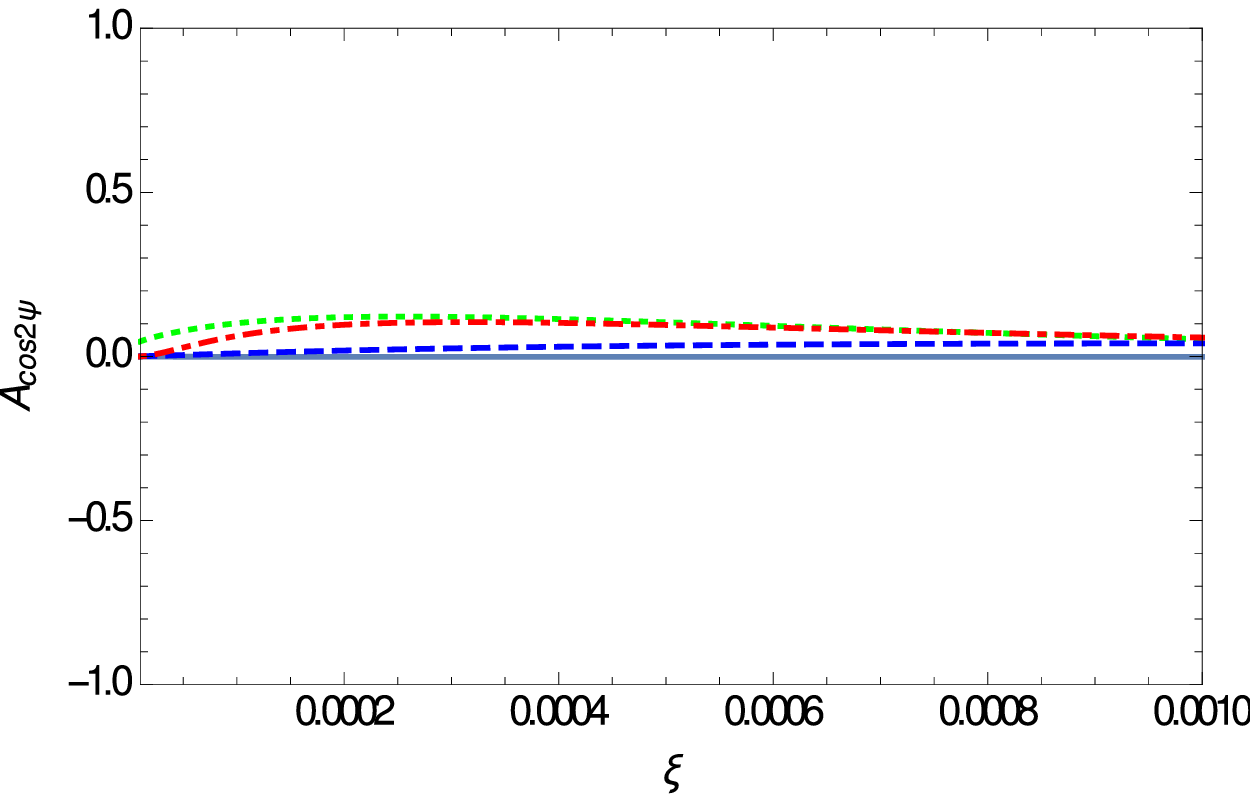}
\includegraphics[scale=0.35]{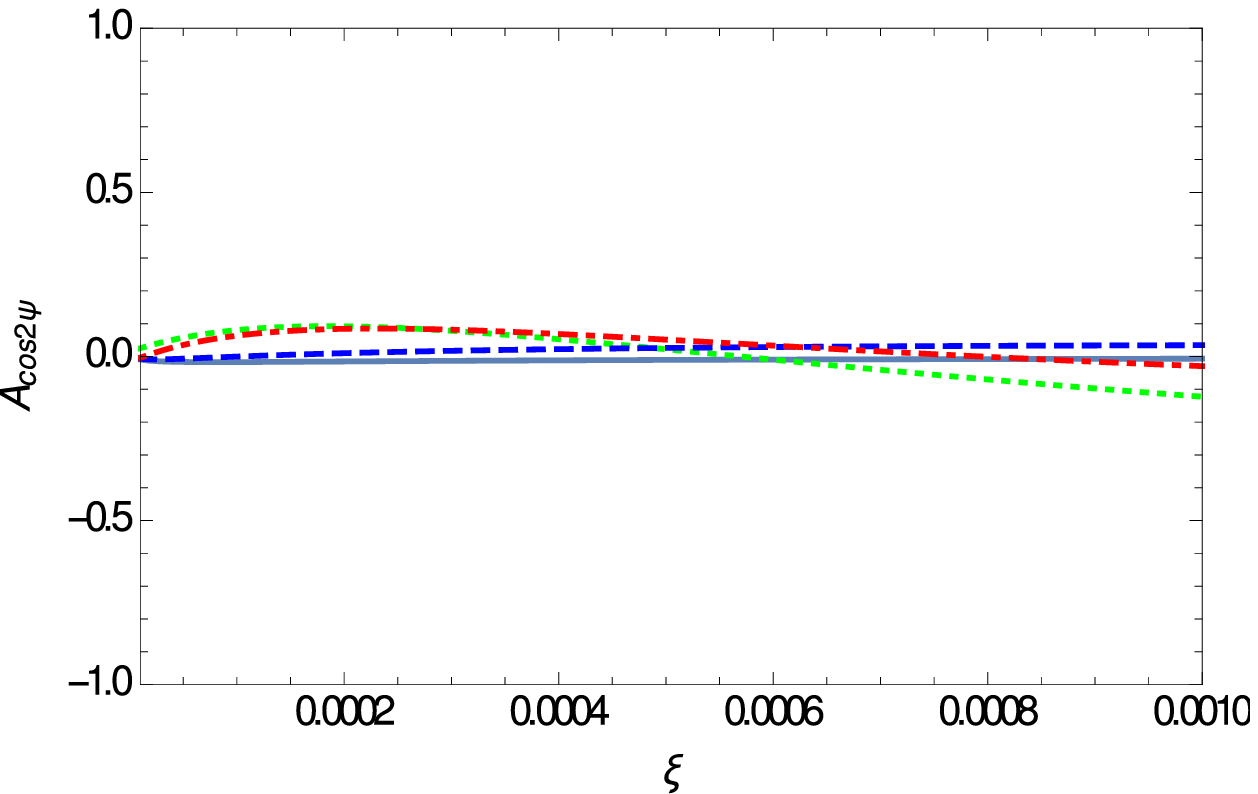}
\includegraphics[scale=0.35]{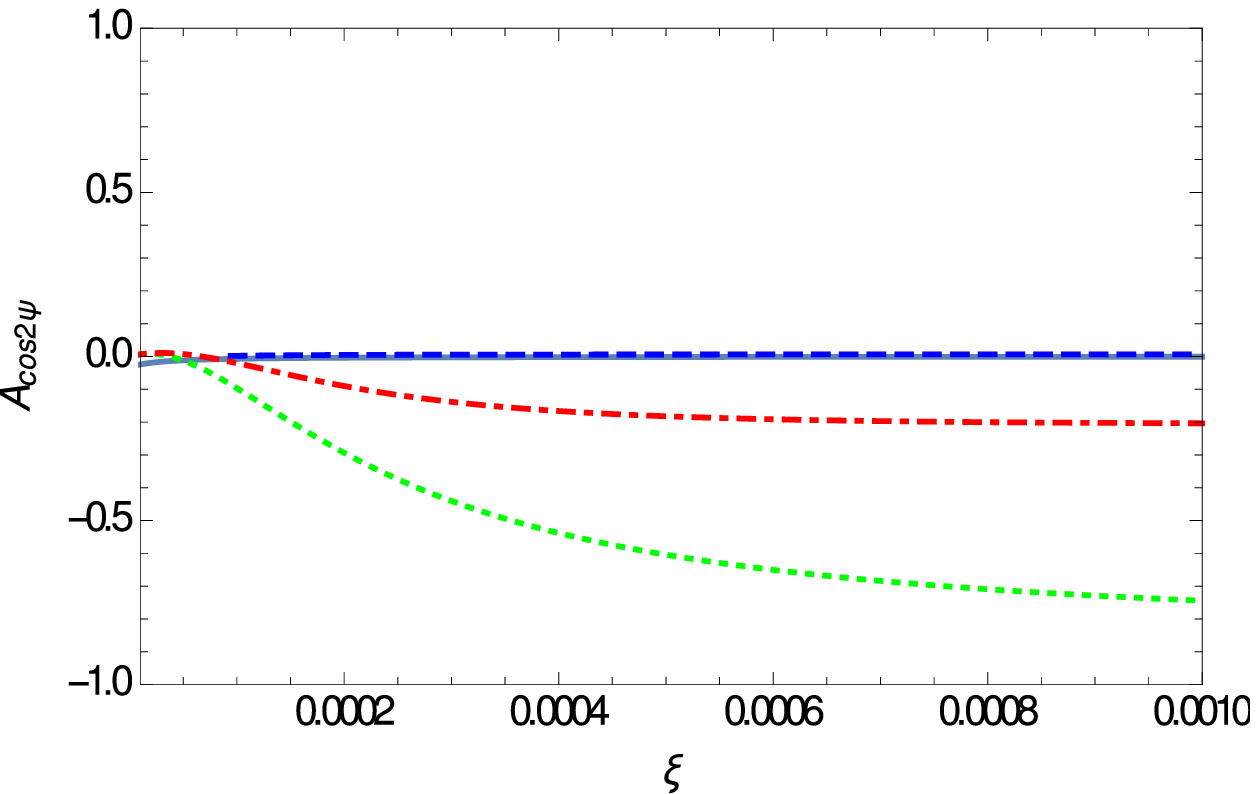}\\
\includegraphics[scale=0.35]{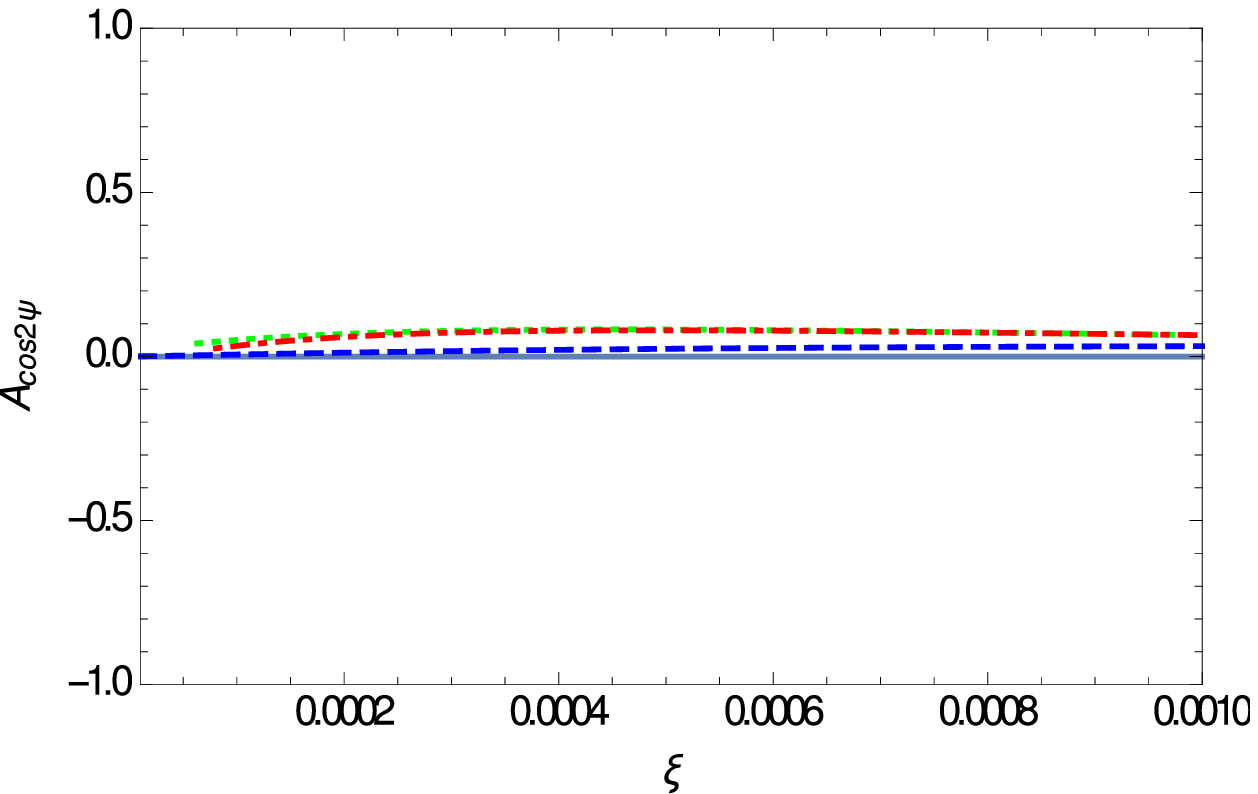}
\includegraphics[scale=0.35]{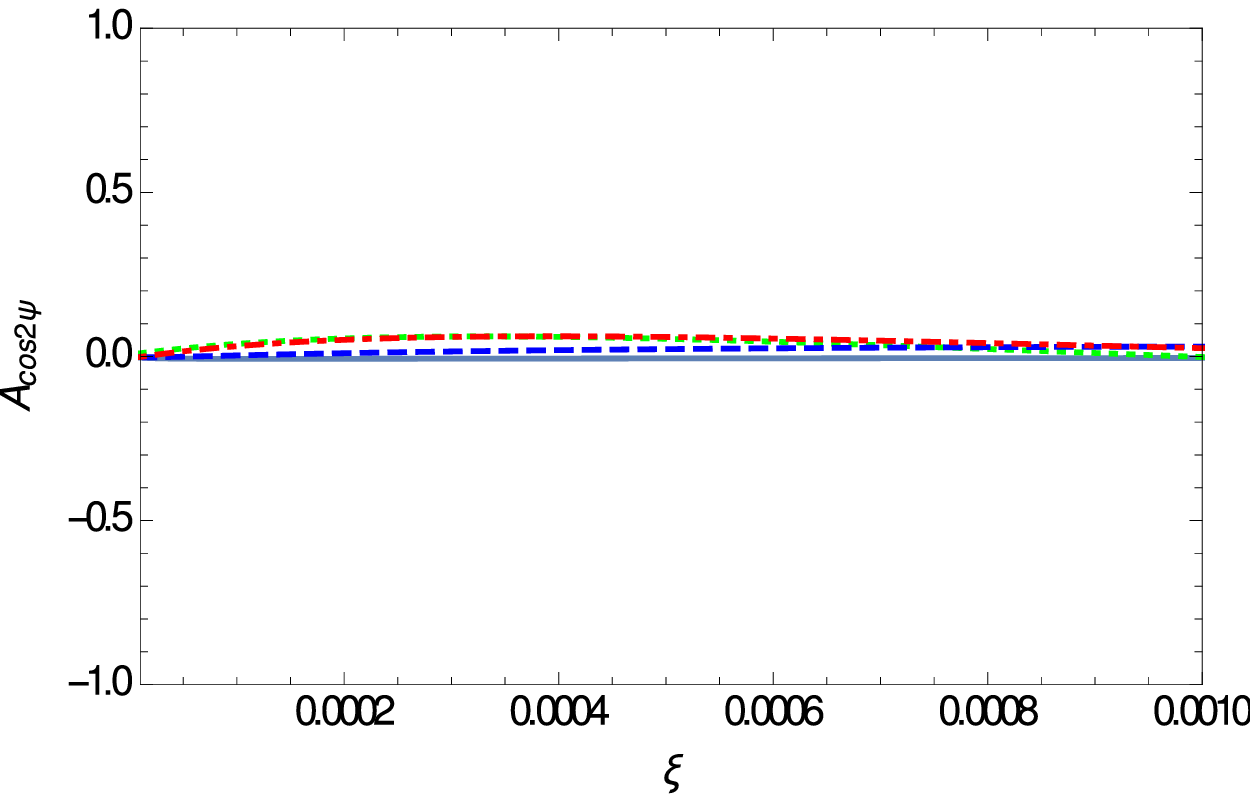}
\includegraphics[scale=0.35]{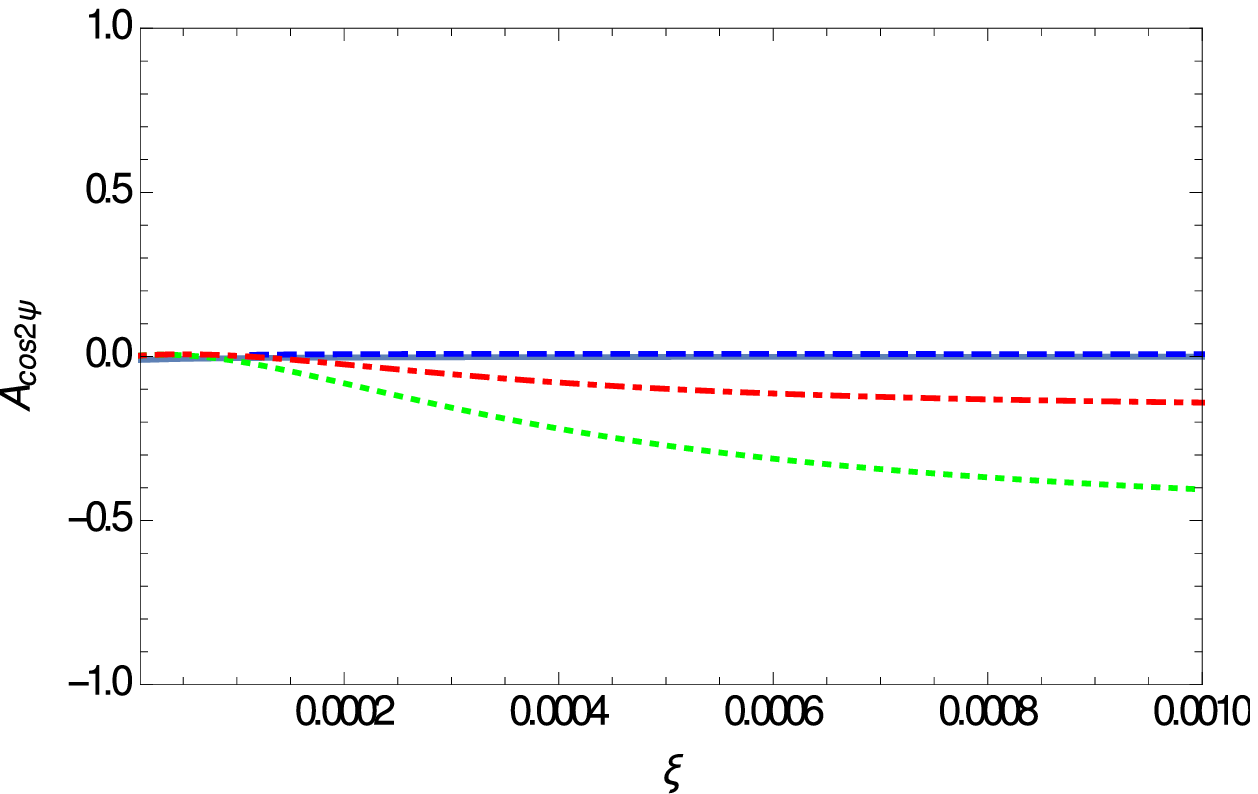}
\caption{\label{fig:cm21}
The $\mathrm{cos}(2\psi)$ modulation as functions of $\xi$ for $x=0.005$.
The upper, mid, and lower rows correspond to $y=0.1$, $0.4$, and $0.7$, respectively,
while the left, mid, and right columns correspond to $z=0.3$, $0.6$, and $0.9$, respectively.
The solid, dotted, dashed, dashdotted curves correspond to the results for the $^3S_1^{[1]}$,
$^1S_0^{[8]}$, $^3S_1^{[8]}$, and $^3P_J^{[8]}$ states, respectively.
}
\end{figure}

\begin{figure}
\includegraphics[scale=0.35]{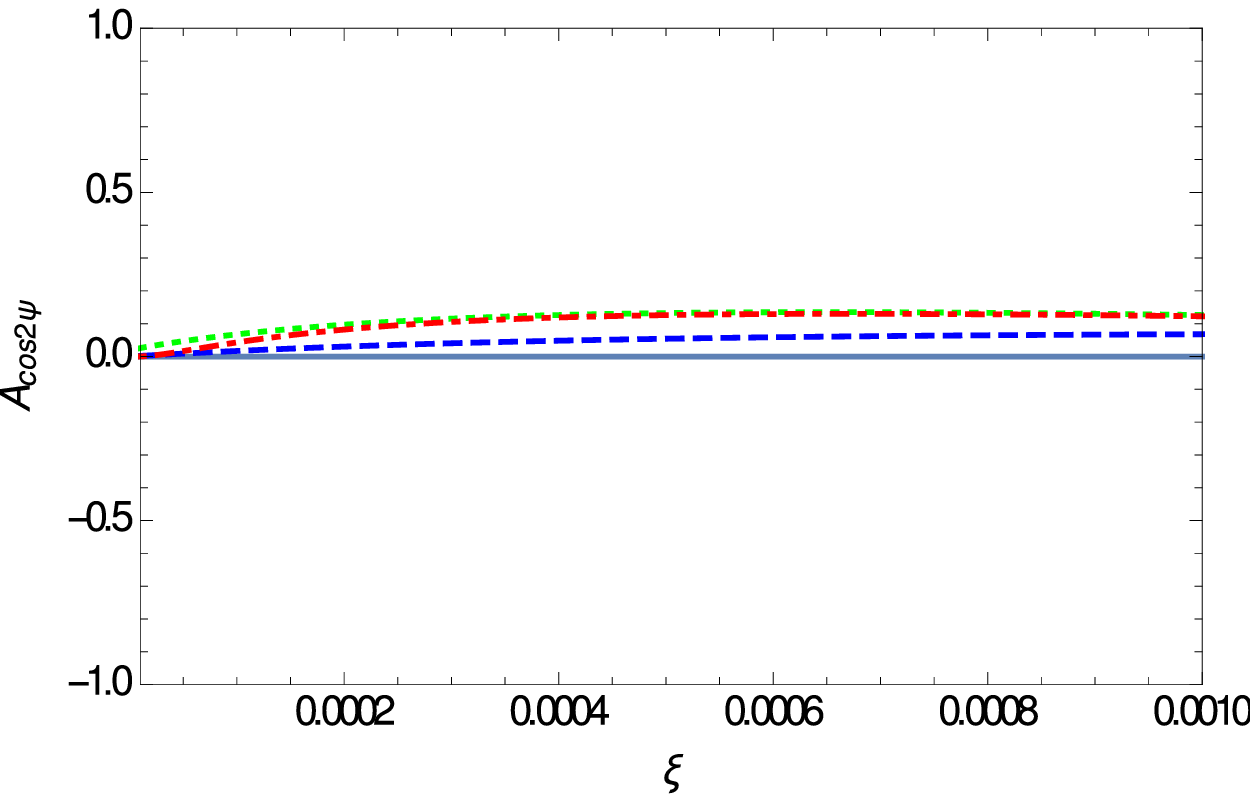}
\includegraphics[scale=0.35]{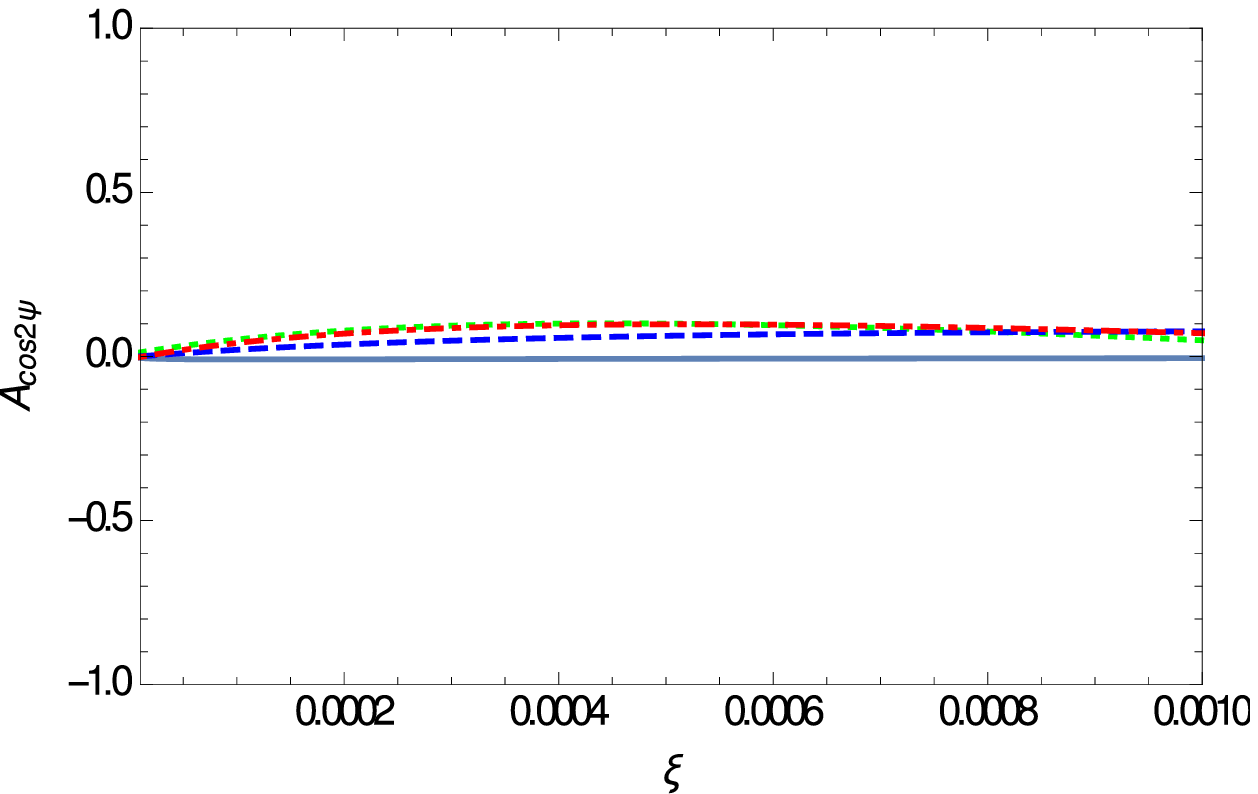}
\includegraphics[scale=0.35]{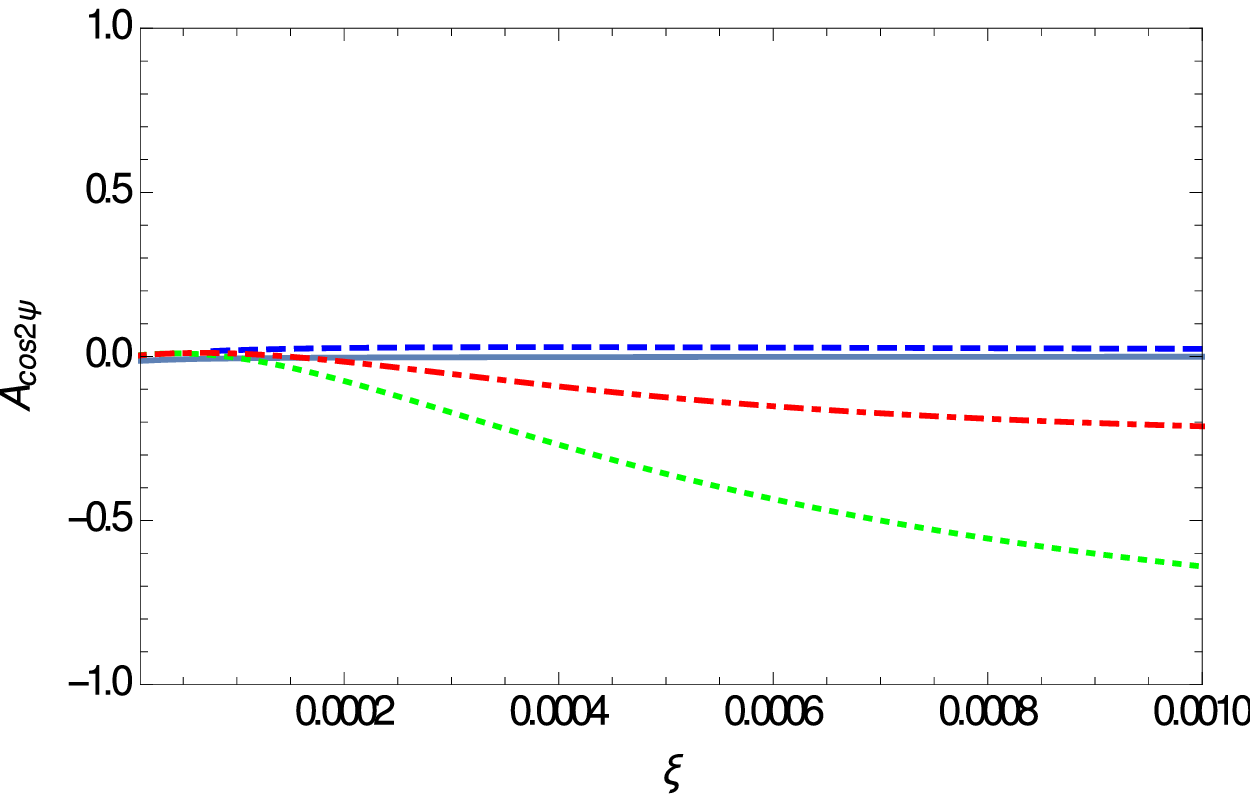}\\
\includegraphics[scale=0.35]{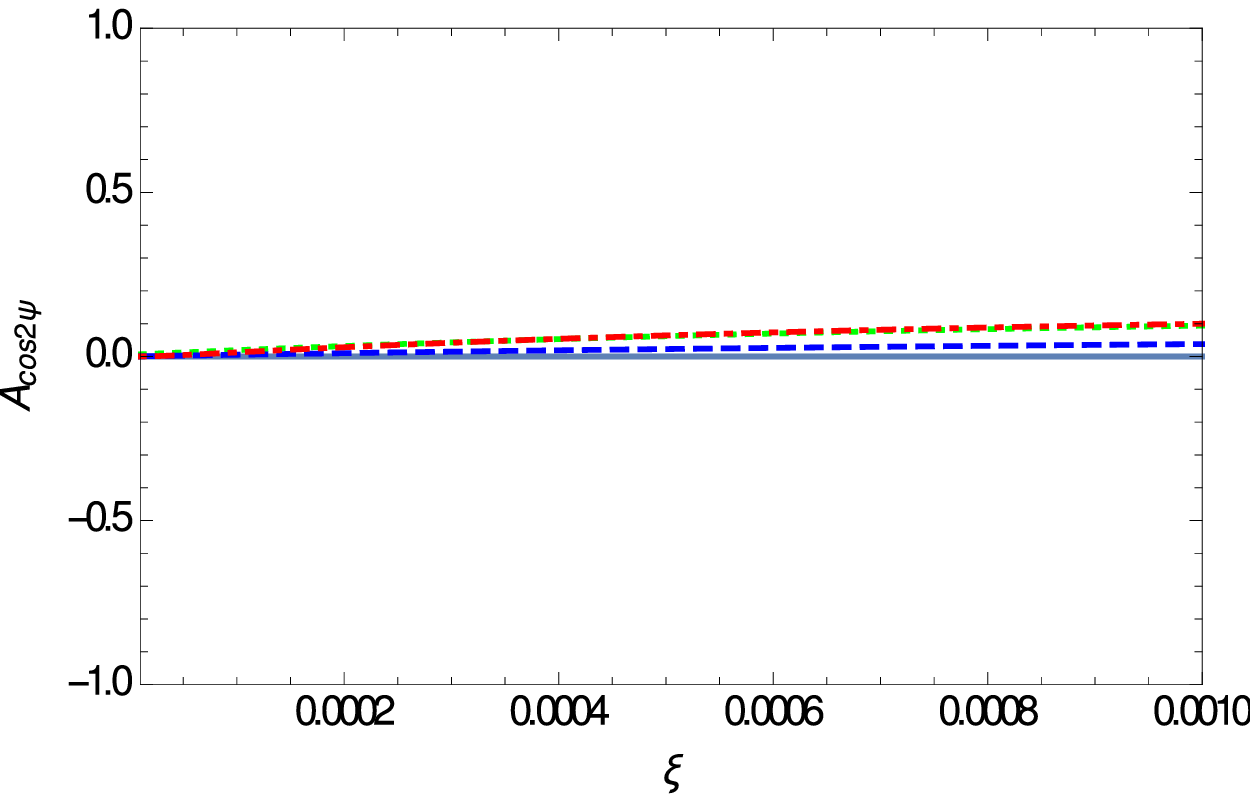}
\includegraphics[scale=0.35]{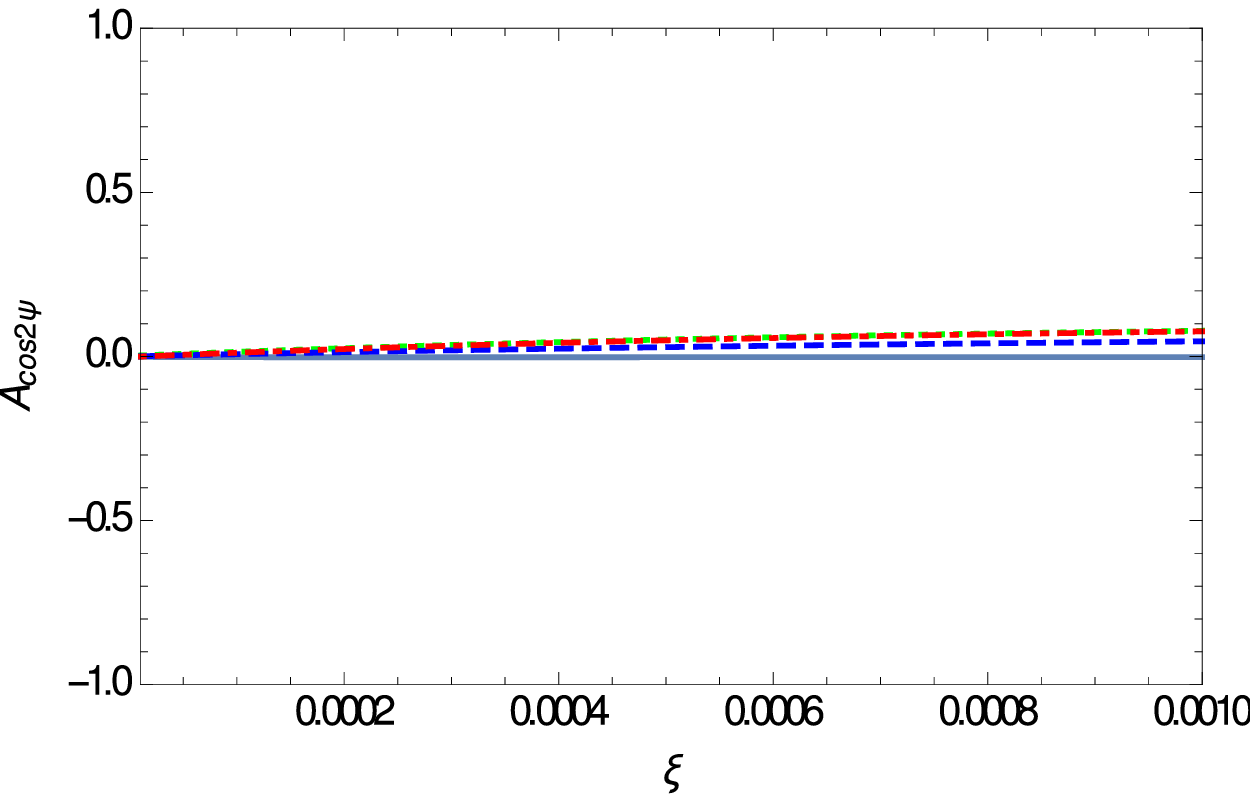}
\includegraphics[scale=0.35]{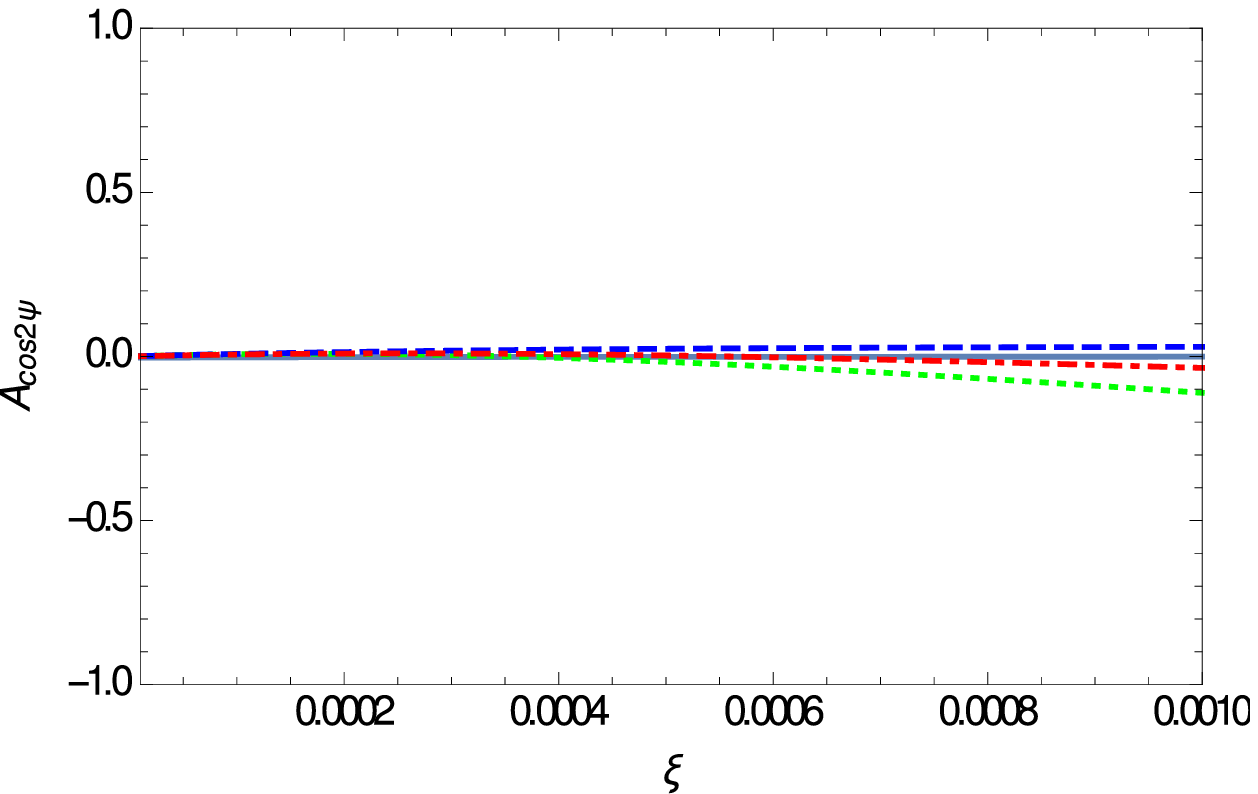}\\
\includegraphics[scale=0.35]{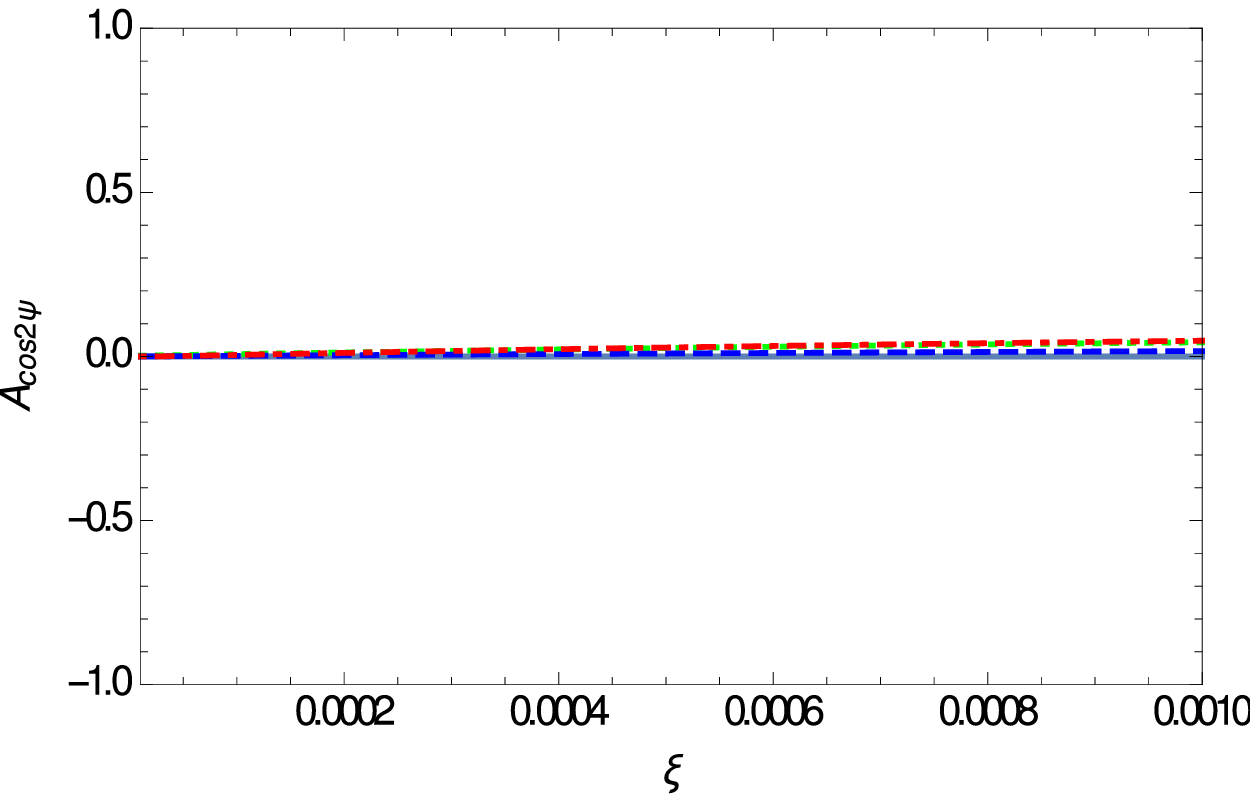}
\includegraphics[scale=0.35]{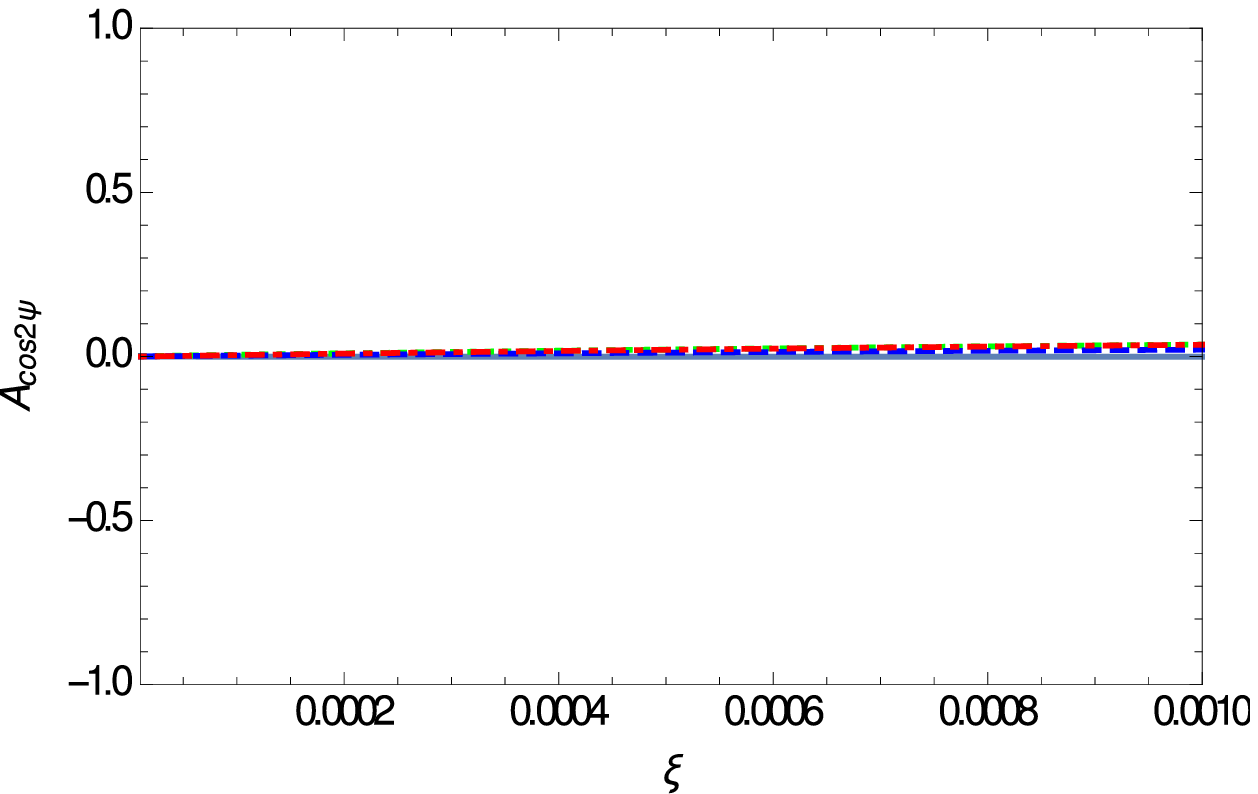}
\includegraphics[scale=0.35]{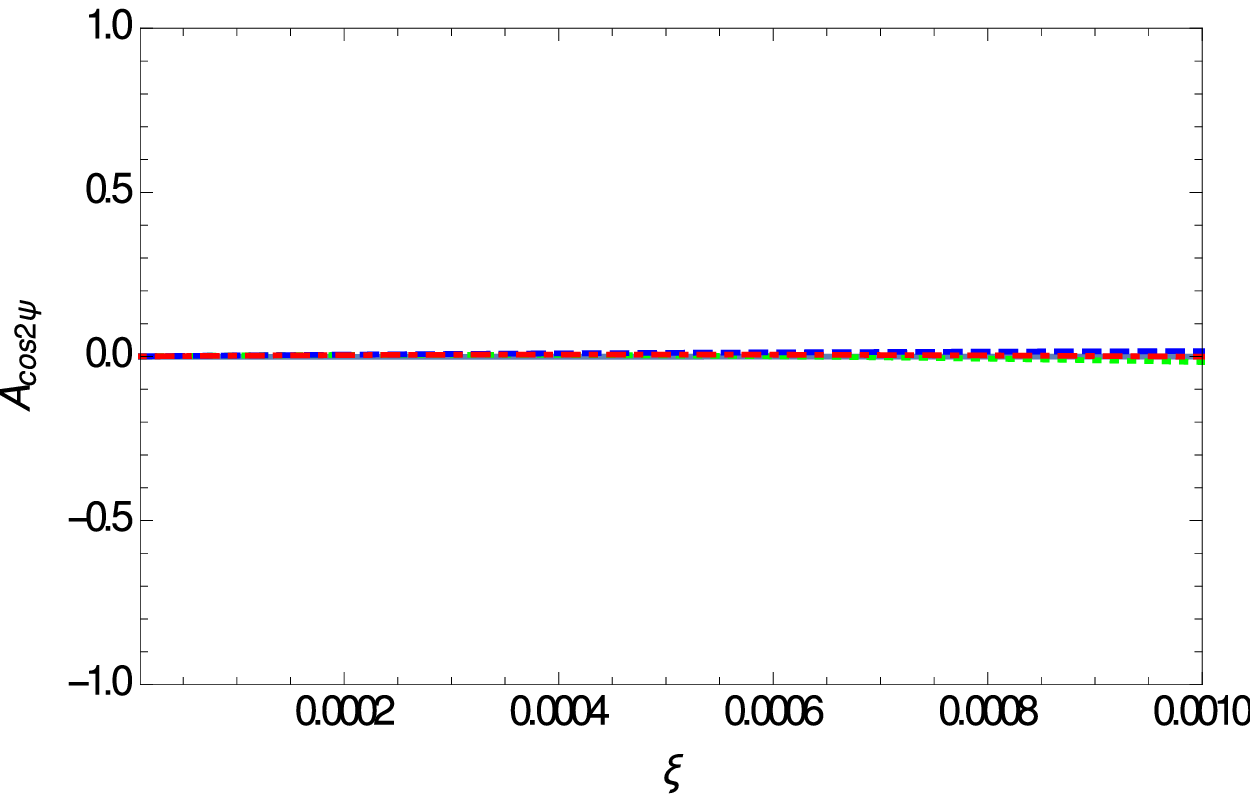}
\caption{\label{fig:cm22}
The $\mathrm{cos}(2\psi)$ modulation as functions of $\xi$ for $x=0.05$.
The upper, mid, and lower rows correspond to $y=0.1$, $0.4$, and $0.7$, respectively,
while the left, mid, and right columns correspond to $z=0.3$, $0.6$, and $0.9$, respectively.
The solid, dotted, dashed, dashdotted curves correspond to the results for the $^3S_1^{[1]}$,
$^1S_0^{[8]}$, $^3S_1^{[8]}$, and $^3P_J^{[8]}$ states, respectively.
}
\end{figure}

\begin{figure}
\includegraphics[scale=0.35]{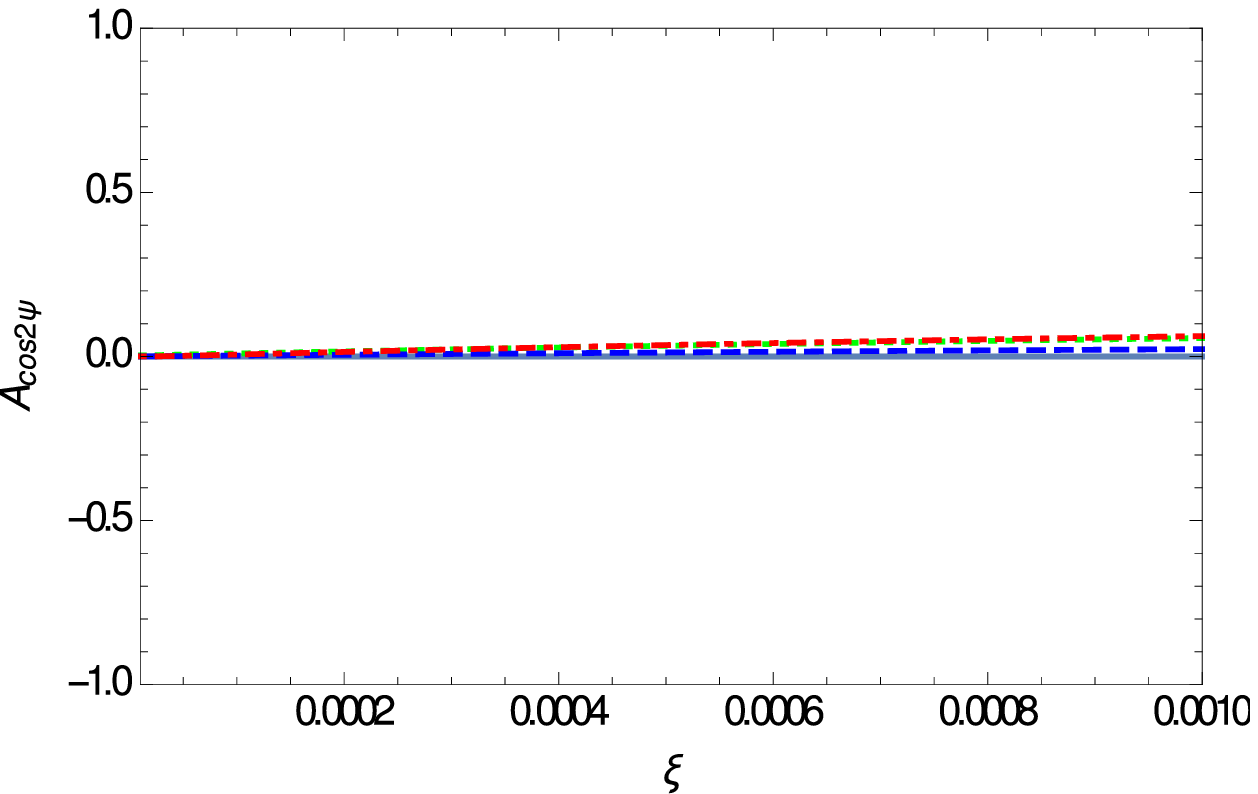}
\includegraphics[scale=0.35]{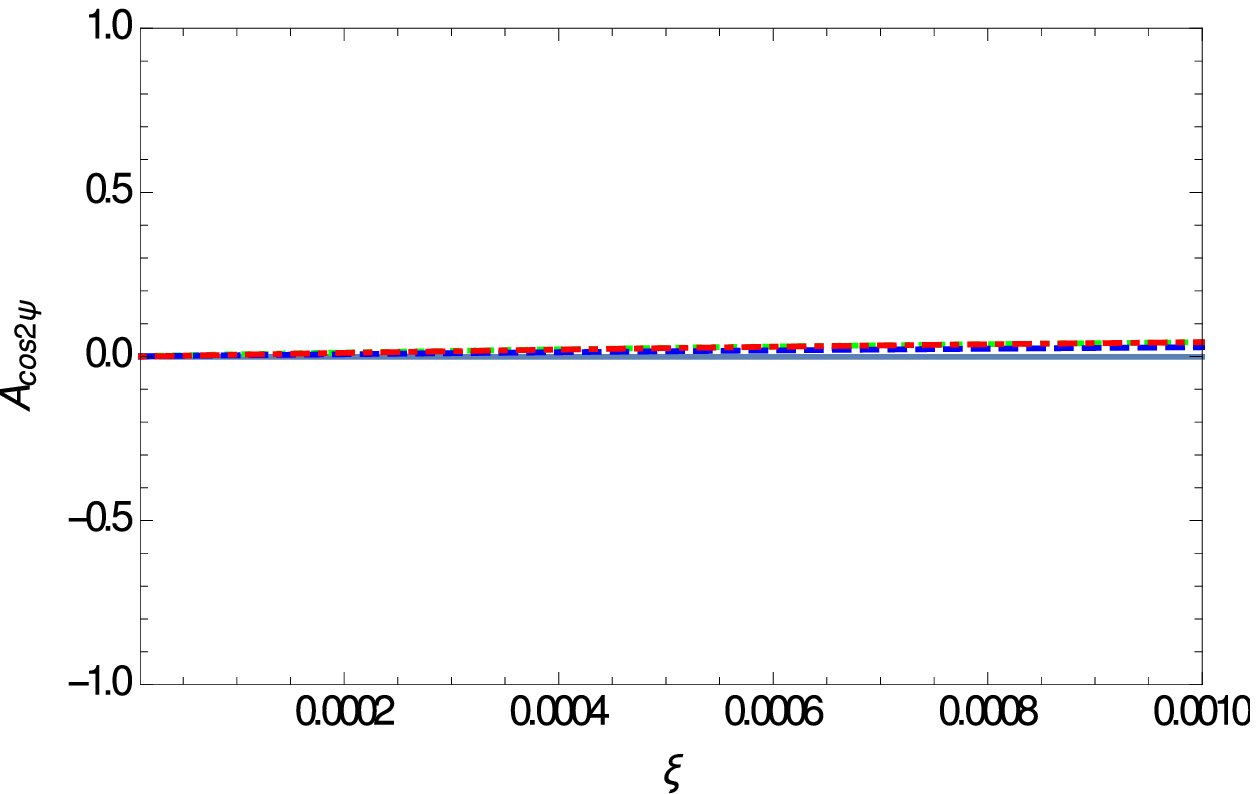}
\includegraphics[scale=0.35]{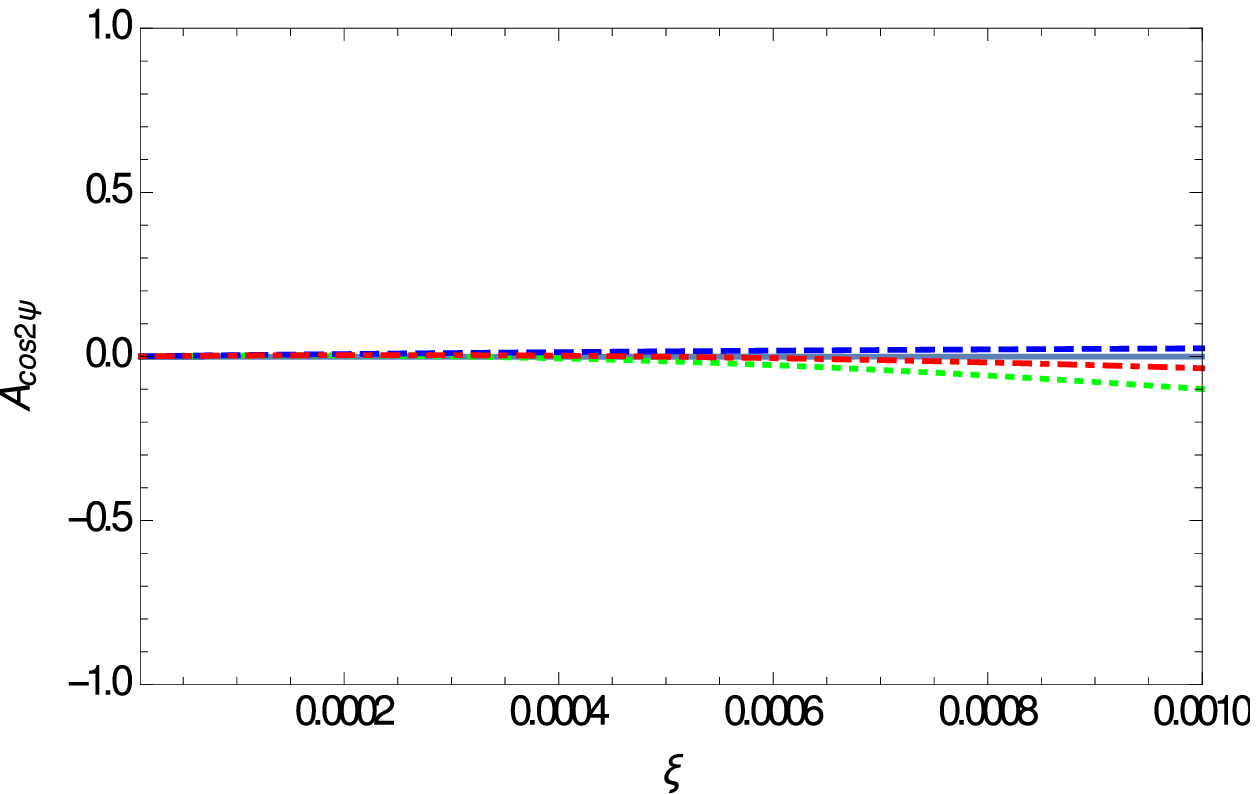}\\
\includegraphics[scale=0.35]{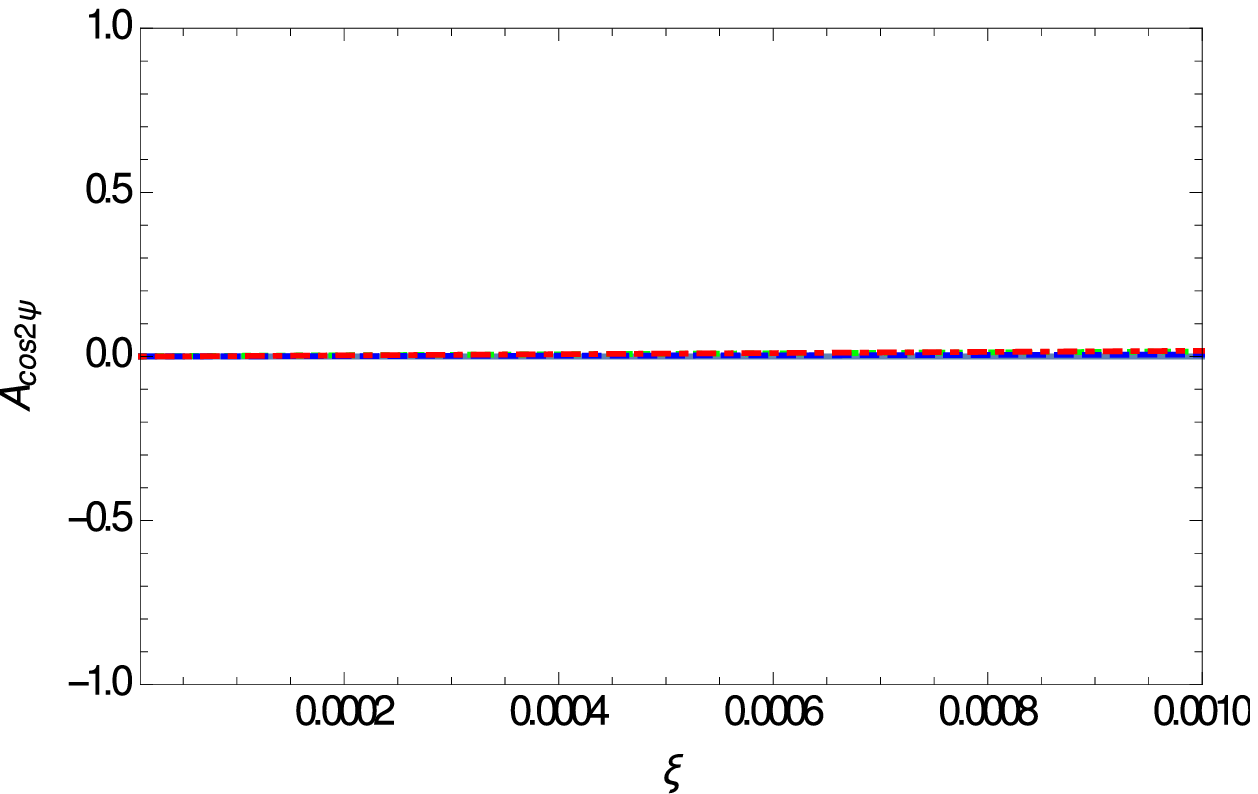}
\includegraphics[scale=0.35]{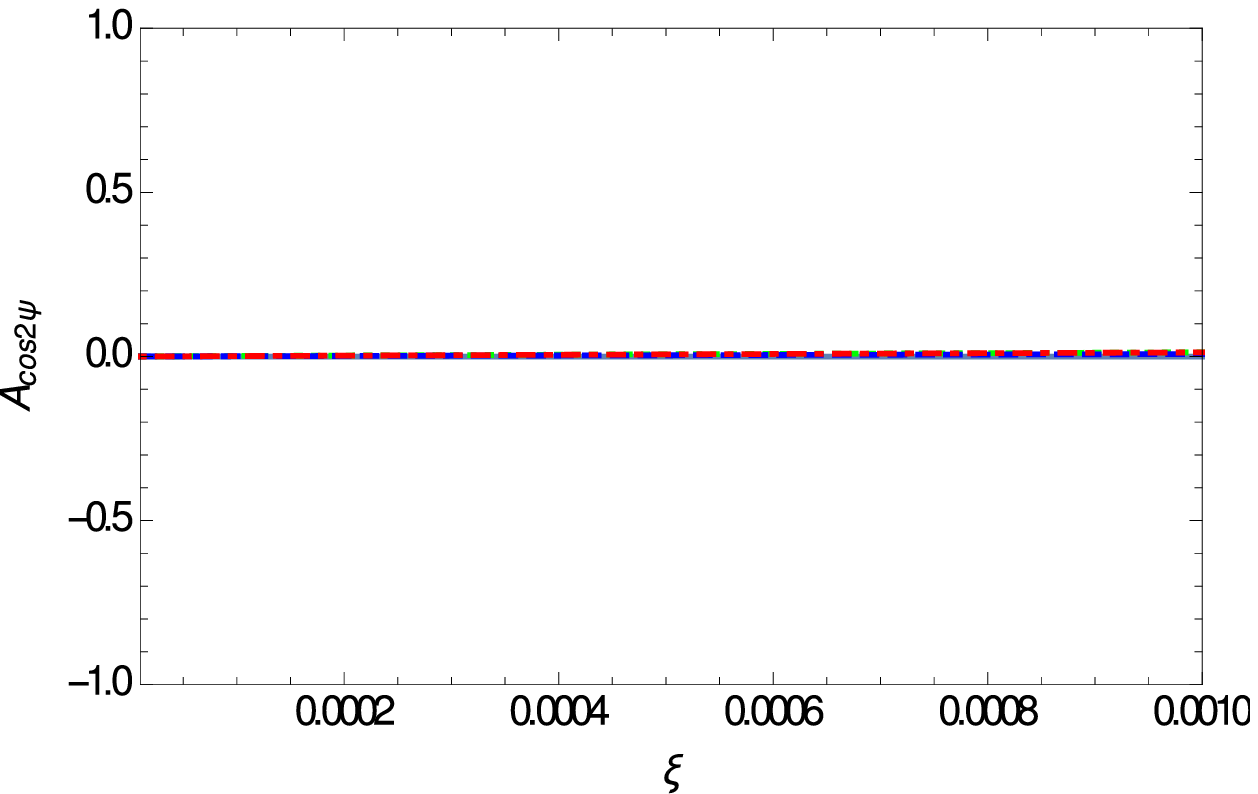}
\includegraphics[scale=0.35]{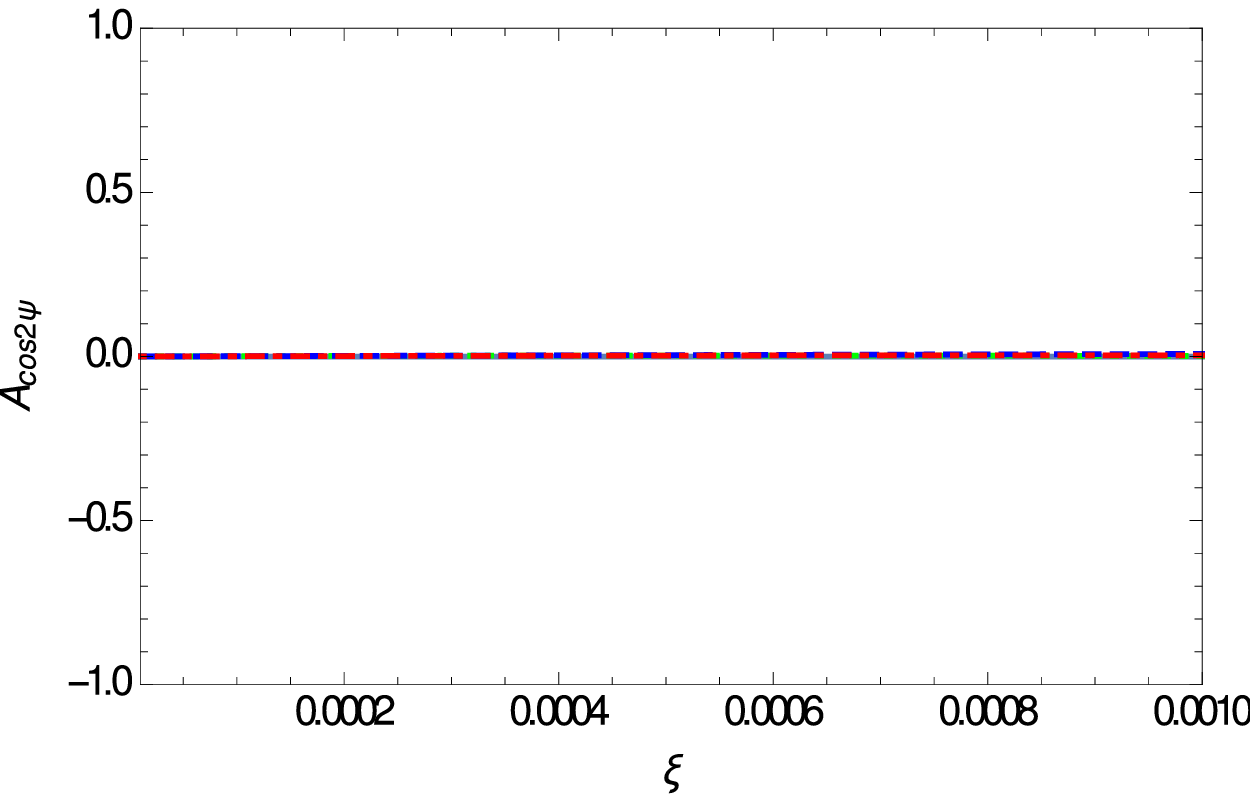}\\
\includegraphics[scale=0.35]{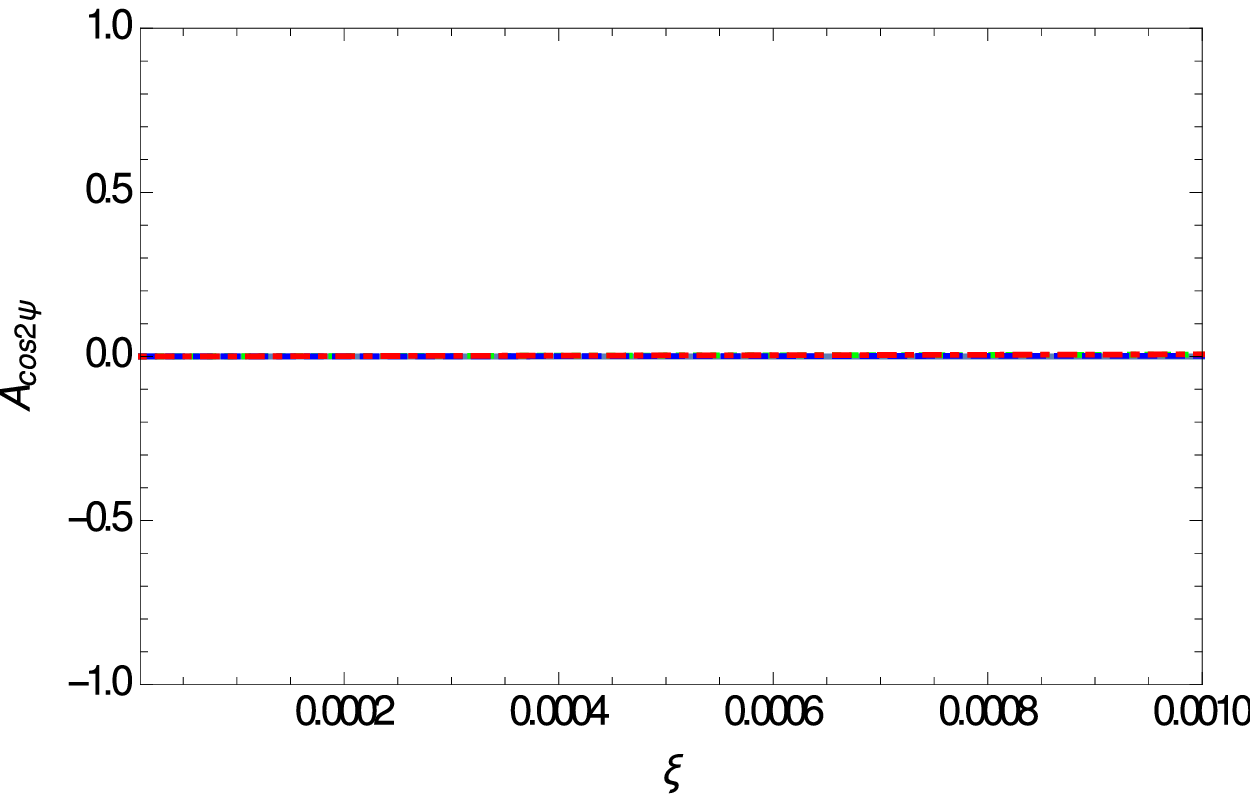}
\includegraphics[scale=0.35]{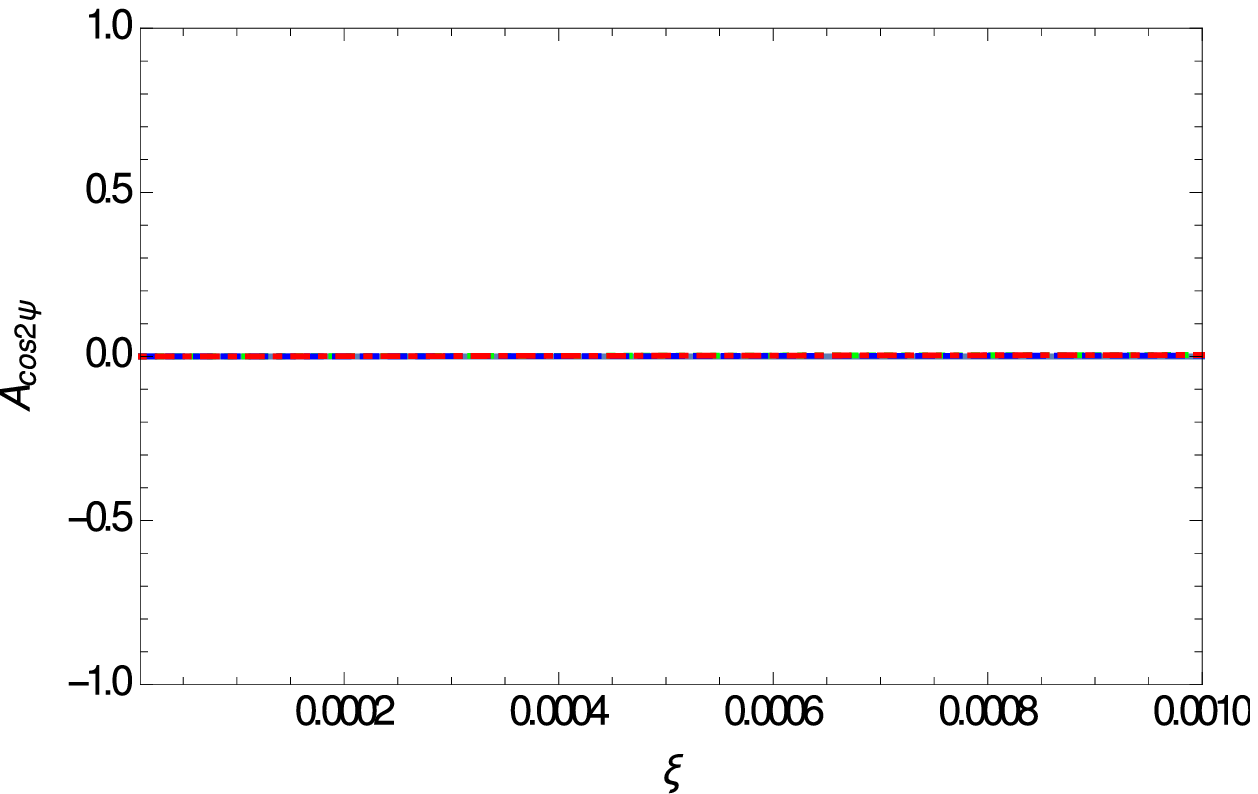}
\includegraphics[scale=0.35]{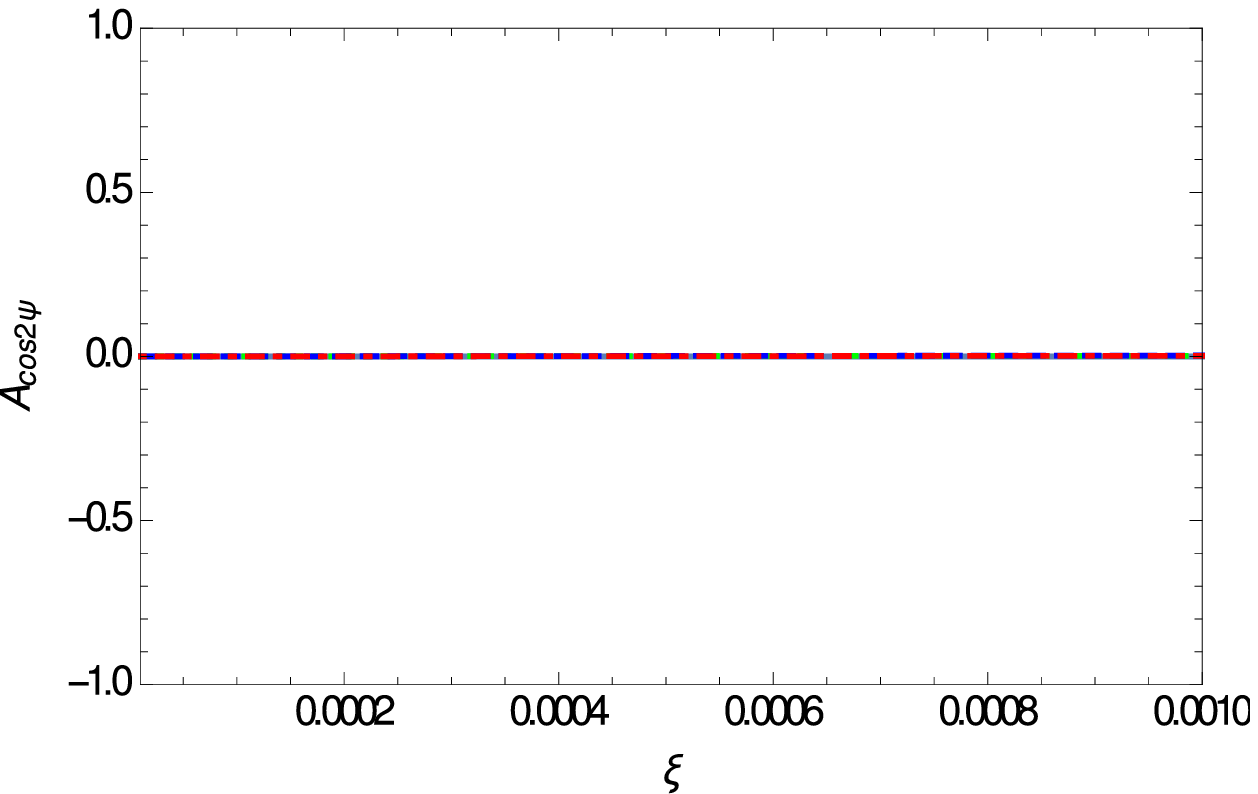}
\caption{\label{fig:cm23}
The $\mathrm{cos}(2\psi)$ modulation as functions of $\xi$ for $x=0.5$.
The upper, mid, and lower rows correspond to $y=0.1$, $0.4$, and $0.7$, respectively,
while the left, mid, and right columns correspond to $z=0.3$, $0.6$, and $0.9$, respectively.
The solid, dotted, dashed, dashdotted curves correspond to the results for the $^3S_1^{[1]}$,
$^1S_0^{[8]}$, $^3S_1^{[8]}$, and $^3P_J^{[8]}$ states, respectively.
}
\end{figure}

With the above parameter choices, we present the values of $A_{\mathrm{cos}\psi}$ in Figure~\ref{fig:cm11}, \ref{fig:cm12}, \ref{fig:cm13},
and those of $A_{\mathrm{cos}2\psi}$ in Figure~\ref{fig:cm21}, \ref{fig:cm22}, \ref{fig:cm23}, for individual $c\bar{c}$ states.
The $\xi$ dependence of the azimuthal asymmetry modulations are presented for any combination of the following parameter choices,
$x=$0.005, 0.05, 0.5, $y=$0.1, 0.4, 0.7, and $z$=0.3, 0.6, 0.9.
The value of $\xi$ ranges from 0.00001 to 0.001,
corresponding to $p_t\approx1\gev$ to $p_t\approx10\gev$.

As an inspiring result,
the values of the azimuthal asymmetry modulations for the four $c\bar{c}$ states are remarkably distinguished in some of these kinematic regions.
To tell the CS and CO channels apart, we can measure $A_{\mathrm{cos}\psi}$ at $x>0.05$, and $z\sim0.9$,
where the values of $A_{\mathrm{cos}\psi}$ for all the three CO states lie around 0,
while those for the CS state is as large as 1.
If, as argued in some papers, the $J/\psi$ production is dominated by the CS channel,
the value of $A_{\mathrm{cos}\psi}$ in this region should coincide with the solid curve presented in the plot at the lower right corner in Figure~\ref{fig:cm13}.
However, this strategy is not feasible because in large-$x$ region,
the cross sections are so small that no enough events can be produced in this region to perform reasonable analysis.
Fortunately, this problem can be amended by including the small-$x$ region,
where although the CS and $^3S_1^{[8]}$ channels cannot be distinguished,
the CS and CO mechanisms are still distinguishable as the $^3S_1^{[8]}$ channel is greatly suppressed in $J/\psi$ leptoproduction,
For this reason, we perform our analysis in the region $x>0.001$ and $0.75<z<0.9$,
in which the value of $Q^2$ is generally larger than $4\gev^2$ and the validity of the perturbative expansion is guaranteed.

Another interesting feature of our numerical results is that the $^1S_0^{[8]}$ channel can also be well distinguished in some kinematic regions.
For $x\sim0.005$, $y\sim0.1$, and $z\sim0.9$, the value of $A_{\mathrm{cos}2\psi}$ for $c\bar{c}(^1S_0^{[8]})$ is almost -1,
while that for all the other three states is almost 0.
Since most of the extractions of the CO LDMEs lead to the same picture,
\textit{i.e.} the $^1S_0^{[8]}$ LDME is at least one order of magnitude greater than the other two,
which however is not consistent with the NRQCD scaling,
this feature provides a perfect laboratory to test the $^1S_0^{[8]}$ dominance picture.
If the $J/\psi$ leptoproduction is also dominated by the $^1S_0^{[8]}$ channel,
we should observe the value of $A_{\mathrm{cos}2\psi}$ at almost -1 in this kinematic region.
Since at larger value of $x$, the cross sections are very small,
we can carry out the analysis in the same kinematic region as in the above paragraph,
namely $x>0.001$ and $0.75<z<0.9$.

In order to draw up a practical strategy for the experimental measurement,
we need to calculate the number of events assuming a luminosity,
and analyse the systematic uncertainties.
For this purpose, we need to specify the value of the renormalization scale $\mu_r$,
the strong coupling constant $\alpha_s$, and electromagnetic fine structure constant $\alpha$.
The value of $\mu_r$ in our calculation is given by $\mu_r=\sqrt{S(xy+\xi)}$.
The running of $\alpha_s$ follows the following equation,
\bea
\alpha_s(\mu_r)=\frac{12\pi}{(33-2n_f)\mathrm{log}(\mu_r^2/\Lambda_{QCD}^2)}, \label{eqn:alphas}
\eea
where at $n_f=5$, $\Lambda_{QCD}$ is given by $\Lambda_{QCD}=0.226\gev$.
It is easy to verify that at the $Z_0$ boson mass $M_Z\approx91\gev$,
the value of $\alpha_s$ is approximately 0.130.
The electromagnetic fine structure constant $\alpha=1/137$ is adopted.

In the following, we carry out our study in the kinematic region,
say $x>0.001$ and $0.75<z<0.9$.
Due to the limit of the capability of the detectors,
we further constrain the value of $y$ and $\xi$ in the region $0.04<y<0.6$ and $\xi>0.00001$.
In the following, we denote this kinematic region as $D$.
This configuration correspond to $Q^2>4\gev^2$, $60\gev<W<240\gev$, and $p_t>1\gev$.
The lower bound of $X$ can be easily obtained as $X_{min}\approx0.0013$,
which is a moderate value to neglect the gluon saturation effects.

Integrating over $x$, $y$, $z$, and $\xi$ in the region $D$,
we obtain the cross sections contributed from the four channels as
\bea
\md\sigma^D\Big|_n=\sigma^D_0\Big|_n
\left[1+A_{\mathrm{cos}\psi}^D\Big|_n\mathrm{cos}(\psi)+A_{\mathrm{cos}2\psi}^D\Big|_n\mathrm{cos}(2\psi)\right]\md\psi, \label{eqn:csd}
\eea
where
\bea
&&\sigma^D_0\Big|_{CS}=5.06\times10^1\times\frac{\langle\mathcal{O}^{J/\psi}(^3S_1^{[1]})\rangle}{M^3}~\mathrm{pb}, \NO \\
&&\sigma^D_0\Big|_{^1S_0^{[8]}}=4.18\times10^3\times\frac{\langle\mathcal{O}^{J/\psi}(^1S_0^{[1]})\rangle}{M^3}~\mathrm{pb}, \NO \\
&&\sigma^D_0\Big|_{^3S_1^{[8]}}=1.08\times10^2\times\frac{\langle\mathcal{O}^{J/\psi}(^3S_1^{[8]})\rangle}{M^3}~\mathrm{pb}, \NO \\
&&\sigma^D_0\Big|_{^3P_J^{[8]}}=6.56\times10^4\times\frac{\langle\mathcal{O}^{J/\psi}(^3P_0^{[8]})\rangle}{M^5}~\mathrm{pb},
\eea
and
\bea
&&A_{\mathrm{cos}\psi}^{D}\Big|_{CS}=0.561,~~~~~~A_{\mathrm{cos}2\psi}^{D}\Big|_{CS}=-0.0658, \NO \\
&&A_{\mathrm{cos}\psi}^{D}\Big|_{^1S_0^{[8]}}=-0.225,~~~~~~A_{\mathrm{cos}2\psi}^{D}\Big|_{^1S_0^{[8]}}=-0.281, \NO \\
&&A_{\mathrm{cos}\psi}^{D}\Big|_{^3S_1^{[8]}}=0.531,~~~~~~A_{\mathrm{cos}2\psi}^{D}\Big|_{^3S_1^{[8]}}=-0.0555, \NO \\
&&A_{\mathrm{cos}\psi}^{D}\Big|_{^3P_J^{[8]}}=-0.346,~~~~~~A_{\mathrm{cos}2\psi}^{D}\Big|_{^3P_J^{[8]}}=0.0220.
\eea

As expected, the CS and CO channels can be well separated by measuring $A_{\mathrm{cos}\psi}^D$,
while the $^1S_0^{[8]}$ channel can be distinguished by studying $A_{\mathrm{cos}2\psi}^D$.
Although the short-distance coefficient for the $^3S_1^{[8]}$ channel is almost twice of that for the CS one,
the cross section for this channel, however,
is negligible due to a much smaller LDME which is suppressed by two orders of magnitude relative to the CS one.
On the experiment side, the $J/\psi$ events are collected in bins of $\psi$.
The azimuthal modulations thus can be extracted by fitting the data to Equation~(\ref{eqn:aac}),
where linear regression is always used as a standard technique.
Since the quantities that we are interested in are ratios,
most of the systematic uncertainties are cancelled when doing the data analysis.
As a result, we consider only the statistical uncertainty in the following.
In our analysis, the range, $[0,~2\pi)$, of $\psi$ is divided into 12 equidistant bins,
namely $[0,~\pi/6)$, $\ldots$, $[11\pi/6,~2\pi)$.
The integral over $\psi$ of the three modulations, 1, $\mathrm{cos}(\psi)$, and $\mathrm{cos}(2\psi)$, in each bin are calculated,
and make up three vectors each of which consists of 12 elements.
Explicitly, they are
\bea
&&v_1=\left(\frac{\pi}{6},~\ldots,~\frac{\pi}{6}\right), \NO \\
&&v_{\mathrm{cos}\psi}=(\frac{1}{2},~-\frac{1}{2}+\frac{\sqrt{3}}{2},~1-\frac{\sqrt{3}}{2},
-1+\frac{\sqrt{3}}{2},~\frac{1}{2}-\frac{\sqrt{3}}{2},~-\frac{1}{2}, \NO \\
&&~~~~~~~~~~-\frac{1}{2},~\frac{1}{2}-\frac{\sqrt{3}}{2},~-1+\frac{\sqrt{3}}{2},
1-\frac{\sqrt{3}}{2},~-\frac{1}{2}+\frac{\sqrt{3}}{2},~\frac{1}{2}), \NO \\
&&v_{\mathrm{cos}2\psi}=\left(\frac{\sqrt{3}}{4},~0,~-\frac{\sqrt{3}}{4},~-\frac{\sqrt{3}}{4},~0,~\frac{\sqrt{3}}{4},
\frac{\sqrt{3}}{4},~0,~-\frac{\sqrt{3}}{4},~-\frac{\sqrt{3}}{4},~0,~\frac{\sqrt{3}}{4}\right).
\eea
It is easy to verify that the three vectors are linearly independent and not correlated.

If the $J/\psi$ leptoproduction is dominated by the CS mechanism,
the integrated cross section in the region $D$ can be expressed as
\bea
\md\sigma^D\Big|_{CS}=2.17\left[1+0.561\mathrm{cos}(\psi)-0.0658\mathrm{cos}(2\psi)\right]\md\psi~\mathrm{pb}, \label{eqn:cscs}
\eea
where
\bea
\langle\mathcal{O}^{J/\psi}(^3S_1^{[1]})\rangle=1.16\gev^3
\eea
is employed~\cite{Eichten:1995ch}.
Assuming the integrated luminosity at the future $ep$ collider is $\mathcal{L}=10^3\mathrm{pb}^{-1}$,
we can construct the similar vector, as in the above paragraph, for this distribution as
\bea
&&v_d=(1683,~1582,~1361,~1035,~691,~466, \NO \\
&&~~~~~~~~466,~691,~1036,~1361,~1582,~1683), \label{eqn:vd}
\eea
each element of which is expectation of the number of events collected in the corresponding bin.
This number, denoted as $v_d^i$, obey Gaussian distribution,
the standard error of which can be obtained as the square root of $v_d^i$.
Denoting the probability density function of the number of events in each bin as $f_i(N_i)$, $i=1,~\ldots,~12$,
we fit $N_i$ to the following equation,
\bea
N_i=\mathcal{A}v_1^i+\mathcal{B}v_{\mathrm{cos}\psi}^i+\mathcal{C}v_{\mathrm{cos}2\psi}^i.
\eea
We can then obtain the expectation values of $\mathcal{A}$, $\mathcal{B}$, and $\mathcal{C}$,
for each combination of the values of $N_i$.
These expectation values thus have a probability density $\prod_if_i(N_i)$.
Then we can calculate the expectation value of the coefficients in Equation~(\ref{eqn:csd}) for each combination of $N_i$,
and the probability density function with respect to $\sigma_0$, $A_{\mathrm{cos}\psi}$, and $A_{\mathrm{cos}2\psi}$.
It is easy to demonstrate that $\mathcal{A}$, $\mathcal{B}$, and $\mathcal{C}$ also obey Gaussian distribution,
and there expectation values and standard errors are
$\overline{\mathcal{A}}=2.17\times10^3$, $\sigma_{\mathcal{A}}=18.6$,
$\overline{\mathcal{B}}=1.22\times10^3$, $\sigma_{\mathcal{B}}=26.2$,
and $\overline{\mathcal{C}}=-0.143\times10^3$, $\sigma_{\mathcal{C}}=27.5$, respectively.
Apparently, the expectation values obtained here agree with Equation~(\ref{eqn:cscs}).
The uncertainties for $A_{\mathrm{cos}\psi}$ and $A_{\mathrm{cos}2\psi}$ can be obtained as
\bea
\Delta A_{\mathrm{cos}\psi}=0.013. \label{eqn:error1}
\eea

If the CO parts are not negligible, we need to sum over their contributions.
Employing the LDMEs obtained in References~\cite{Zhang:2014ybe, Sun:2015pia},
which are given below,
\bea
&&\langle\mathcal{O}^{J/\psi}(^3S_1^{[1]})\rangle=0.645\gev^3, \NO \\
&&\langle\mathcal{O}^{J/\psi}(^1S_0^{[1]})\rangle=0.785\times10^{-2}\gev^3, \NO \\
&&\langle\mathcal{O}^{J/\psi}(^3S_1^{[8]})\rangle=1.0\times10^{-2}\gev^3, \NO \\
&&\langle\mathcal{O}^{J/\psi}(^3P_J^{[1]})\rangle=3.8\times10^{-2}\gev^5, \label{eqn:LDME1}
\eea
the cross section can be expressed as
\bea
\md\sigma^D\Big|_{NRQCD}=12.7\left[1-0.246\mathrm{cos}(\psi)-0.0155\mathrm{cos}(2\psi)\right].
\eea
This equation leads to $A_{\mathrm{cos}\psi}=-0.246$.
It is easy to see that this number is well separated from that led to by the CS model in the sense of the uncertainty given in Equation~(\ref{eqn:error1}).

In the rest of this paper,
we will demonstrate that the azimuthal asymmetry in the $J/\psi$ leptoproduction
can also distinguish the LDMEs that are consistent with the $^1S_0^{[8]}$ dominance picture.
Taking those given in Reference~\cite{Chao:2012iv} as an example, say
\bea
&&\langle\mathcal{O}^{J/\psi}(^3S_1^{[1]})\rangle=1.16\gev^3, \NO \\
&&\langle\mathcal{O}^{J/\psi}(^1S_0^{[1]})\rangle=8.9\times10^{-2}\gev^3, \NO \\
&&\langle\mathcal{O}^{J/\psi}(^3S_1^{[8]})\rangle=0.3\times10^{-2}\gev^3, \NO \\
&&\langle\mathcal{O}^{J/\psi}(^3P_J^{[1]})\rangle=1.26\times10^{-2}\gev^5, \label{eqn:LDME2}
\eea
The cross section for the $J/\psi$ production in the region $D$ can be expressed as
\bea
\md\sigma^D\Big|_{\mathrm{Chao}}=19.4\left[1-0.158\mathrm{cos}(\psi)-0.203\mathrm{cos}(2\psi)\right]\md\psi~\mathrm{pb}.
\eea
Using the same analysis strategy, we can obtain
\bea
A_{\mathrm{cos}2\psi}\Big|_{\mathrm{Chao}}=-0.203\pm0.004.
\eea
Since with the LDMEs given in Equation~(\ref{eqn:LDME1}),
the value of $A_{\mathrm{cos}2\psi}$ is -0.0155,
it is obvious that one can distinguish the two sets of LDMEs by scrutinize the value of $A_{\mathrm{cos}2\psi}$.

\section{Summary\label{sec:summary}}

In this paper, we calculated the azimuthal asymmetry modulations in the $J/\psi$ leptoproduction,
namely $A_{\mathrm{cos}\psi}$ and $A_{\mathrm{cos}2\psi}$ defined in Equation~(\ref{eqn:adef}),
as functions of $x$, $y$, $z$, and $\xi$.
By scrutinizing their behaviours,
we find that the CS and CO mechanisms can be well distinguished through the measurement of the values of $A_{\mathrm{cos}\psi}$.
Further, the $^1S_0^{[8]}$ dominance picture can also be tested by measuring the values of $A_{\mathrm{cos}2\psi}$.
Restricted in the region, $0.001<x<1$, $0.04<y<0.6$, $0.75<z<0.9$, and $p_t>1\gev$,
we carried out the calculation of the differential cross sections with respect to the azimuthal angle for three models,
and found that they lead to clearly different values of the azimuthal-asymmetry modulations.
Having applied rigorous statistical analysis, we found that at the integrated luminosity $\mathcal{L}=1000\mathrm{pb}^{-1}$,
the statistical uncertainties of $A_{\mathrm{cos}\psi}$ and $A_{\mathrm{cos}2\psi}$ are small enough to tell the three models apart.
As a conclusion, the azimuthal asymmetry in the $J/\psi$ leptoproduction provides a good laboratory for the study of the quarkonium production mechanisms.
We suggest that this experiment be implemented at the future $ep$ colliders such as the EIC.

\acknowledgments

We thank Zhan Sun for his contribution at the early stage.
This work is supported by the National Natural Science Foundation of China (Grant Nos. 11605144, 11747037).
Y.-P. Y. acknowledges support from SUT and the Office of the Higher Education Commission under the National Research Universities project of Thailand.

%\bibliography{paper}% Produces the bibliography via BibTeX.
\providecommand{\href}[2]{#2}\begingroup\raggedright\endgroup

\end{document}